\documentstyle[preprint,eqsecnum,aps]{revtex}
\tightenlines

\begin{document}
\draft
\title{Magnetohydrodynamics in the Early Universe and the Damping of Non-linear
Alfv\'en Waves }
\author{Kandaswamy Subramanian$^*$ and John D. Barrow}
\address{Astronomy Centre, University of Sussex, Brighton BN1 9QH, UK.}
\date{\today }
\maketitle

\begin{abstract}
The evolution and viscous damping of cosmic magnetic fields in the early
universe, is analysed. Using the fact that the fluid, electromagnetic, and
shear viscous energy-momentum tensors are all conformally invariant, the
evolution is transformed from the expanding universe setting into that in
flat spacetime. Particular attention is paid to the evolution of nonlinear
Alfv\'en modes. For a small enough magnetic field, which satisfies our
observational constraints, these wave modes either oscillate negligibly or,
when they do oscillate, become overdamped. Hence they do not suffer Silk
damping on galactic and subgalactic scales. The smallest scale which
survives damping depends on the field strength and is of order a
dimensionless Alfv\'en velocity times the usual baryon-photon Silk damping
scale. After recombination, nonlinear effects can convert the Alfv\'en mode
into compressional, gravitationally unstable waves and seed cosmic
structures if the cosmic magnetic field is sufficiently strong.
\end{abstract}

\preprint{HEP/123-qed}


\narrowtext

\section{Introduction \protect\\}

\label{sec:level1}

The origin of ordered, large-scale cosmic magnetic fields remains a
challenging problem. It is widely assumed that magnetic fields in
astronomical objects, like galaxies, grew by turbulent dynamo amplification
of small seed magnetic fields \cite{dynam}. There have been many suggestions
for producing the small seed magnetic fields required to prime the
amplification process \cite{seedrev}. Many of these proposals appeal to
processes which might operate in the very early universe \cite{euseed}.
Moreover, the efficiency with which known turbulent dynamo mechanisms can
produce the observed fields is still being debated \cite{dyndeb}. An
alternate possibility is that the galactic field could be the remnant of a
larger cosmological field of primordial origin \cite{primhyp}, although, as
yet, there is no compelling mechanism for producing the required field. It
could form part of the initial conditions, arise at a phase transition, or
be produced in some way at the end of a period of inflation \cite{euseed}.
If the primordial field is homogeneous then the isotropy of the microwave
background places a limit of $3.4\times 10^{-9}(\Omega _0h_{50}^2)^{\frac 12}
$ Gauss on its present strength \cite{barrow}. This limits the strength of
any primordial field on scales greater than the horizon at last scattering.

Dissipative processes play a key role in fashioning the spectrum of
irregularities that survive the radiation era. It is well known that density
fluctuations in the baryons suffer ''Silk'' damping in the early universe,
due to radiative viscosity \cite{silk}. This leads one to wonder whether
fluctuating magnetic fields produced in the early universe, which couple to
the baryons, will also be damped by similar processes. This problem was
considered by Jedamzik, Katalinic and Olinto \cite{JKO96}, who examined the
damping of linearised magnetohydrodynamical (MHD) wave modes in an expanding
universe. However, there remain subtleties about the formulation of the
problem, and the damping processes, which prompt us to reanalyse the problem
in a completely different fashion that also allows non-linear effects to be
considered.

We are able to do this by exploiting the conformal invariance of the
relativistic fluid, electromagnetic, and shear viscous energy-momentum
tensors to transform the problem from the expanding universe setting to a
simpler problem in flat spacetime. This transformation is explained in
detail in the next section and in Appendix A, and it allows us to study the
evolution and damping of various MHD waves. After taking the
non-relativistic limit of the resulting MHD equations, in Section III, we
focus in Section IV on non-linear Alfv\'en wave solutions and study their
damping. The compressible MHD modes are also briefly treated, in their
linearised limit, in Section V. As the universe expands, the relevant
mean-free-path of the particle responsible for the damping eventually
exceeds the wavelength of a given mode. The evolution of various MHD motions
in this free-streaming regime, is considered in Section VI. Eventually, the
universe recombines to form atoms and the tangles in the magnetic field that
survive damping, can create gravitationally unstable perturbations. This
post-recombination regime is considered in Section VII. In Section VIII, we
connect these ideas together to discuss how a given spectrum of magnetic
inhomogeneities evolves, and section IX summarizes our conclusions.

\section{ Conformal invariance of relativistic, Viscous MHD}

We shall make use of a conformal invariance property of the field equations
of general relativity. Consider two metrics which are related by conformal
transformations of the form 
\begin{equation}
g_{\mu \nu }^{*}=\Omega ^2g_{\mu \nu }.  \label{conform}
\end{equation}
where $\Omega \equiv \Omega (x^\alpha ).$ Suppose the cosmic fluid is
described by an energy-momentum tensor, say $T^{\mu \nu }$. If this is
traceless, $T_\mu ^\mu =0$, then the equations of motion of the fluid are
invariant under conformal transformations. That is, we have 
\begin{equation}
T_{\quad ;\nu }^{\mu \nu }=0\quad \Leftrightarrow \quad T_{\quad ;*\nu
}^{*\mu \nu }=0\quad {\rm with}\quad T^{*\mu \nu }=\Omega ^{-6}T^{\mu \nu }.
\label{eqmot}
\end{equation}
Here $;*$ denotes a covariant derivative with respect to the metric $g_{\mu
\nu }^{*}$. This result is straightforward to demonstrate by direct
calculation.

We apply this result to the early universe setting. Firstly, note that the
Friedmann-Robertson-Walker (FRW) metric is conformally flat. We will only
need to consider the zero-curvature metric with line element 
\begin{equation}
ds^2=g_{\mu \nu }dx^\mu dx^\nu =a^2(\tau )\left[ -d\tau
^2+dx^2+dy^2+dz^2\right] ,  \label{metric}
\end{equation}
where $\tau $ is conformal time and $a(\tau )$ is the scale factor. We also
define the comoving proper time $t$ using $dt=ad\tau $. We can relate this
metric to flat spacetime, $\eta _{\mu \nu },$ by a conformal transformation, 
$\eta _{\mu \nu }=g_{\mu \nu }^{*}=a^{-2}(\tau )g_{\mu \nu }$, with $\Omega
=a^{-1}(\tau )$. So if the energy-momentum tensor of the matter in the early
universe is trace-free and transforms under conformal transformations as in
Eq. (\ref{eqmot}), then we can transform the fluid equation of motion to a
simpler equation in flat spacetime. This approximation method allows us to
transform non-linear solutions to the MHD equations, for a fluid with
trace-free energy-momentum tensor, from flat spacetime into an expanding FRW
universe. We are, of course, neglecting the back-reaction of those motions
upon the form of the metric. That is, the gravitational potential (metric)
perturbations due to the inhomogeneous motions must be small for this
approximation to hold. For example, if the inhomogeneous motions of a
radiation fluid give rise to energy density inhomogeneities of amplitude $%
\delta \rho /\rho $ over a length scale $\lambda ,$ then the metric
perturbations induced are of order $(\delta \rho /\rho )(\lambda /ct)^2$.
These are small because, for the parameters we consider, $\delta \rho /\rho
<<1$ and the perturbations are on sub-horizon scales.

The energy-momentum tensor in the early universe will be modelled by a
linear combination of ideal fluid, non-ideal fluid, and electromagnetic
fields. We have 
\begin{equation}
T^{\mu \nu }=T_I^{\mu \nu }+T_{NI}^{\mu \nu }+T_{EM}^{\mu \nu },
\label{eten}
\end{equation}
where we have separated the energy-momentum tensor into an ideal fluid part $%
T_I^{\mu \nu }$, a non-ideal fluid part $T_{NI}^{\mu \nu }$ and an
electromagnetic part $T_{EM}^{\mu \nu }$. We take the ideal part to be a
perfect fluid with equation of state $p=\rho /3$, where $p$ is the pressure
and $\rho $ is the energy density (we have set the speed of light $c\equiv 1$%
.) For most of the period before decoupling, the early universe is radiation
dominated and one can use the above equation of state. Later, we will
comment on the effects of matter-domination ($p=0$) before decoupling. Thus,
we have 
\begin{equation}
T_I^{\mu \nu }=(p+\rho )U^\mu U^\nu +pg^{\mu \nu }.  \label{fluidet}
\end{equation}
where $U^\mu $ is the normalised four-velocity of the plasma, satisfying the
condition $U^\mu U_\mu =-1$. Note that for $p=\rho /3$, we have $T_{I\mu
}^\mu =0$. Under a conformal transformation, the components of $T_I^{\mu \nu
}$ transform to a new set of (starred) variables as follows 
\begin{equation}
p^{*}=\Omega ^{-4}p,\quad \rho ^{*}=\Omega ^{-4}\rho ,\quad U^{*\mu }=\Omega
^{-1}U^\mu  \label{tranfl}
\end{equation}
and the ideal fluid energy-momentum tensor transforms as $T_I^{*\mu \nu
}=\Omega ^{-6}T_I^{\mu \nu }$.

The non-ideal fluid part of the energy-momentum tensor can be written as 
\cite{weinb} 
\begin{equation}
T_{NI}^{\mu \nu }=-\eta H^{\mu \alpha }H^{\nu \beta }W_{\alpha \beta },
\label{nonidet}
\end{equation}
where 
\begin{equation}
H^{\alpha \beta }\equiv g^{\alpha \beta }+U^\alpha U^\beta ;\quad W_{\alpha
\beta }\equiv U_{\alpha ;\beta }+U_{\beta ;\alpha }-{\frac 23}g_{\alpha
\beta }U_{\quad ;\gamma }^\gamma  \label{nonidef}
\end{equation}
and $\eta $ is the effective shear viscosity coefficient, given by (see
Weinberg \cite{weinb}); 
\begin{equation}
\eta ={\frac 4{15}}g{\frac{\pi ^2}{30}}T^4l_d,  \label{etabef}
\end{equation}
where $T$ is the temperature of the radiation-dominated universe, $g$ the
statistical weight and $l_d$ the mean-free-path of the diffusing particle.
In the radiation-dominated epoch, the bulk viscosity is zero and we have
neglected the thermal conductivity term since it does not affect the
Alfv\'en wave mode that we will focus upon.

One can easily check that the trace $T_{NI\mu }^\mu =0$. Also, under
conformal transformation, using $U^{*\alpha }=\Omega ^{-1}U^\alpha $ and $%
U_\alpha ^{*}=\Omega U_\alpha $, we have 
\begin{equation}
W_{\alpha \beta }^{*}=\Omega W_{\alpha \beta }-\left[ U_\alpha {\frac{%
\partial \Omega }{\partial x^\beta }}+U_\beta {\frac{\partial \Omega }{%
\partial x^\alpha }}\right]  \label{wtr}
\end{equation}
and $H^{*\mu \alpha }H^{*\nu \beta }=\Omega ^{-4}H^{\mu \alpha }H^{\nu \beta
}$; so the non-ideal stress conformally transforms as 
\begin{equation}
T_{NI}^{*\mu \nu }=-\eta ^{*}\Omega ^{-3}H^{\mu \alpha }H^{\nu \beta
}W_{\alpha \beta }=\Omega ^{-6}T_{NI}^{\mu \nu },\quad {\rm if}\quad \eta
^{*}\equiv \Omega ^{-3}\eta .  \label{tranonid}
\end{equation}

The electromagnetic part of the energy-momentum tensor is given by 
\begin{equation}
T_{EM}^{\mu \nu }={\frac 1{4\pi }}\left[ F_\gamma ^\mu F^{\nu \gamma }-{%
\frac 14}g^{\mu \nu }F_{\gamma \delta }F^{\gamma \delta }\right] ,
\label{emet}
\end{equation}
where $F_{\mu \nu }\equiv A_{\nu ;\mu }-A_{\mu ;\nu }=A_{\nu ,\mu }-A_{\mu
,\nu }$ is the Maxwell tensor, $A_\mu $ the four-potential, and a comma
denotes an ordinary derivative. This is also traceless: $T_{EM\mu }^\mu =0$.
The evolution of the electromagnetic field is governed by the Maxwell
equations 
\begin{equation}
F_{\quad ;\nu }^{\mu \nu }=4\pi J^\mu ,\quad F_{[\mu \nu ,\gamma ]}=0.
\label{maxwell}
\end{equation}
Here, $J^\mu $ is the four-current density. Under a conformal
transformation, these equations are also invariant provided we transform the
electromagnetic field tensor and current density as follows: 
\begin{equation}
F^{*\mu \nu }=\Omega ^{-4}F^{\mu \nu };\quad J^{*\mu }=\Omega ^{-4}J^\mu .
\label{tranem}
\end{equation}
Under these transformations we verify that 
\begin{equation}
T_{EM}^{*\mu \nu }=\Omega ^{-6}T_{EM}^{\mu \nu }.  \label{eltran}
\end{equation}
Finally, consider the relativistic generalisation of Ohm's law, 
\begin{equation}
J^\mu +J^\nu U_\nu U^\mu =\sigma F^{\mu \nu }U_\nu ,  \label{ohm}
\end{equation}
where $\sigma $ is the conductivity of the fluid, measured in the fluid rest
frame. Transforming the four-current and velocity to the starred frame, we
can see that Ohm's law also remains invariant under conformal
transformations if we define a starred conductivity, $\sigma ^{*}=\Omega
^{-1}\sigma $.

>From these preliminaries, we see that the total energy-momentum tensor
consisting of an ideal fluid, non-ideal fluid, and electromagnetic field is
trace-free, and also transforms as in Eq. (\ref{eqmot}) under a conformal
transformation. Suppose we choose $\Omega =a^{-1}(\tau )$, with the scale
factor normalised such that $a(\tau _0)=1$ at present conformal time $\tau
_0 $. Define a new set of starred variables as in Eqs. (\ref{tranfl}), (\ref
{tranonid}) and (\ref{tranem}). The evolution equations of the starred
variables then become,

\begin{eqnarray}
\  &&T_{\quad ,\nu }^{*\mu \nu }=T_{I\quad ,\nu }^{*\mu \nu }+T_{NI\quad
,\nu }^{*\mu \nu }+T_{EM\quad ,\nu }^{*\mu \nu }=0  \nonumber \\
\  &&F_{\quad ,\nu }^{*\mu \nu }=4\pi J^{*\mu },\quad F_{[\mu \nu ,\gamma
]}^{*}=0  \nonumber \\
\  &&J^{*\mu }+J^{*\nu }U_\nu ^{*}U^{*\mu }=\sigma ^{*}F^{*\mu \nu }U_\nu
^{*}  \label{flateq}
\end{eqnarray}
Since these equations are simply the flat spacetime equations of
relativistic, viscous MHD, we can carry over all results obtained in the
flat spacetime context to MHD in FRW spacetime on scales such that the
metric perturbations created by the magnetic inhomogeneities and their
resulting motions are small. The only extra complication introduced by
cosmology is the time-variability of the viscosity and conductivity
coefficients. (The transformation of MHD equations from the FRW universe to
flat space, has also been explicitly constructed in ref. (\cite{brand}).
However, they did not include the viscous stress, nor exploit the simple
conformal invariance properties stressed here. The idea that flat space
solutions can be taken over to FRW universe in the radiation era, has also
been used by Liang \cite{laing}, to study shock waves in the radiation era;
but, without the inclusion of magnetic fields).

At this stage, it is useful to point out that any particle species which
does not interact with the relativistic fluid is also easy to take into
account. Such a non-interacting component may be any species of dark matter,
or neutrinos, after they have decoupled from the rest of the matter. The
energy-momentum tensor for this component will then be separately conserved.
As long as this component perturbs the form of the FRW metric negligibly,
the above treatment of conformal transformation of relativistic MHD to flat
spacetime will go through. This also means that the MHD equations in the
above form can be applied to the baryon-photon-magnetic field system, even
after dark matter domination, until the baryon density becomes comparable to
the radiation density, which happens close to decoupling. In the sections to
follow, we will use these equations to examine the evolution of MHD waves,
in particular the non-linear Alfv\'en mode, after taking the
non-relativistic limit of the equations.

\section{ Non-relativistic limit}

Suppose the universe was seeded with magnetic fields at some early epoch in
its radiation-dominated phase. We consider the subsequent evolution. For
magnetic fields of realistic strengths (ie that are not going to produce
highly discordant microwave background fluctuations or radically change the
course of primordial nucleosynthesis), the induced bulk fluid velocities are
in general highly non-relativistic. Suppose we write the four-velocity in
the transformed metric as $U^{*\mu }\equiv (\gamma ,\gamma {\bf v})$, where $%
\gamma \equiv (1-{\bf v}^2)^{-1/2}$. This form satisfies the normalisation
condition on the four-velocity and ${\bf v}=d{\bf x}/d\tau =ad{\bf x}/dt$ is
exactly the peculiar bulk velocity of the fluid in the FRW metric. The
non-relativistic limit corresponds to taking $|{\bf v}|<<1$ in the fluid
conservation equation in Eq. (\ref{flateq}) .

If we define the electric field ${\bf E}^{*}\equiv (E^{*1},E^{*2},E^{*3})$
and the magnetic field ${\bf B}^{*}\equiv (B^{*1},B^{*2},B^{*3})$ in the
starred metric by 
\begin{equation}
F^{*0i}=E^{*i}\quad F^{*12}=B^{*3}\quad F^{*23}=B^{*1}\quad F^{*31}=B^{*2},
\label{elbdef}
\end{equation}
then the time component of the fluid equation in Eq. (\ref{flateq}), in the
non-relativistic limit, is 
\begin{equation}
{\frac{\partial \rho ^{*}}{\partial \tau }}+{\bf \nabla }.[(\rho ^{*}+p^{*})%
{\bf v}]-{\bf E}^{*}.{\bf J}^{*}-\eta ^{*}{\bf \nabla }.{\bf f}=0.
\label{energ}
\end{equation}
Here, ${\bf J}^{*}$ is the current-density vector whose components are the
spatial components of the transformed four-current $J^{*\mu }$, and ${\bf f}=%
{\bf \nabla }({\bf v}^2/2)-(2/3){\bf v}{\bf \nabla }.{\bf v}$. The spatial
components give the Euler equation 
\begin{eqnarray}
{\frac \partial {\partial \tau }}\left[ (\rho ^{*}+p^{*}){\bf v}\right] +(%
{\bf v}.{\bf \nabla })\left[ (\rho ^{*}+p^{*}){\bf v}\right] &+&{\bf v}{\bf %
\nabla }.\left[ (\rho ^{*}+p^{*}){\bf v}\right]  \nonumber \\
\ &=&-{\bf \nabla }p^{*}+{\bf J}^{*}\times {\bf B}^{*}+\eta ^{*}\left[
\nabla ^2{\bf v}+{\frac 13}{\bf \nabla }({\bf \nabla }.{\bf v})\right] .
\label{euler}
\end{eqnarray}
We have assumed here that the net charge density ($J^{*0}$) is negligible.

The Maxwell equations in the starred metric, are 
\begin{equation}
{\bf \nabla }\times {\bf B}^{*}=4\pi {\bf J}^{*}+{\frac{\partial {\bf E}^{*}%
}{\partial \tau}}\quad {\bf \nabla }\times {\bf E}^{*}=-{\frac{\partial {\bf %
B}^{*}}{\partial \tau}}\quad {\bf \nabla }.{\bf B}^{*}=0\quad {\bf \nabla }.%
{\bf E}^{*}=4\pi J^{*0}  \label{maxstar}
\end{equation}
They are supplemented by the non-relativistic limit of Ohm's law 
\begin{equation}
{\bf E}^{*}+{\bf v}\times {\bf B}^{*}={\frac{{\bf J}^{*}}{\sigma ^{*}}.}
\label{ohmstar}
\end{equation}
The Maxwell equations in a "Lab frame", with fields ${\bf E}$ and ${\bf B}$
defined using the fundamental observers of FRW spacetime, are also easy to
write down and are given in Appendix A.

In what follows we shall make the usual assumption that, for
non-relativistic velocities, the displacement current term can be neglected.
In this case, we can split up the Lorentz force term in the canonical way as 
${\bf J}^{*}\times {\bf B}^{*}=-{\bf \nabla }{\bf B}^{*2}/(8\pi )+({\bf B}%
^{*}.{\bf \nabla })({\bf B}^{*}/4\pi )$. We also assume that the early
universe was a perfect conductor (see for example ref\cite{euseed}) and take
the limit of $\sigma \propto \sigma ^{*}\to \infty $, in Ohm's law. The
magnetic field then satisfies the ideal limit of the induction equation 
\begin{equation}
{\frac{\partial {\bf B}^{*}}{\partial \tau }}={\bf \nabla }\times \left[ 
{\bf v}\times {\bf B}^{*}\right] .  \label{induct}
\end{equation}
In terms of the ''Lab'' magnetic field, defined in Appendix A, the
ideal-limit induction equation is 
\begin{equation}
{\frac{\partial (a^2{\bf B})}{\partial t}}={\frac 1a}{\frac{\partial (a^2%
{\bf B})}{\partial \tau }}={\frac 1a}{\bf \nabla }\times \left[ {\bf v}%
\times (a^2{\bf B})\right] .  \label{ind}
\end{equation}
Let us turn now to consider solutions of the above equations.

Consider first the unperturbed evolution, with a zero peculiar velocity and
negligible magnetic fields. The solution of the fluid equations is then $%
p^{*}=\rho ^{*}/3=C_1$ with $C_1$ a positive constant. This implies that, in
the original variables, $p=\rho /3=C_1/a^4$, as expected for the
radiation-dominated universe. (Furthermore, for zero peculiar velocity, for
any ''test'' magnetic field, ${\bf B}^{*}$ is constant in time, or ${\bf B}%
\propto a^{-2}$, a result which is intuitively expected for the ''Lab''
magnetic field due to flux freezing in the expanding universe.) Now consider
the effect of introducing tangled magnetic fields in the universe. The
Lorentz force associated with a tangled field will cause the fluid to move
and induce a non-zero peculiar velocity. The coupled system of equations
describing the evolution of the velocity and magnetic fields is highly
non-linear. For this reason, the authors of ref. \cite{JKO96} examined only
the case of weak perturbations around a quasi-uniform field. We shall to
begin with, follow a complementary approach and look at special non-linear
solutions. This will also give some feel for how a general tangled field
configuration will evolve.

\section{Non-linear Alfv\'en waves in the early universe}

\subsection{ The ideal non-viscous regime}

At sufficiently early times, (or, equivalently, for the field and velocity
on sufficiently large scales), one may assume the matter is a perfect fluid
and neglect any viscous effects. Also, the fluid radiation pressure in the
early universe is typically much larger than the magnetic pressure, for the
field strengths we are considering. Their ratio is given by 
\begin{equation}
{\frac{B^{*2}}{8\pi p^{*}}}={\frac{B^2}{8\pi p}}\approx 3\times
10^{-7}B_{-9}^2.  \label{bpres}
\end{equation}
where $B_{-9}$ is the present-day magnetic field in units of $10^{-9}$
Gauss. Here, we have assumed that the fields are simply frozen into the
uniformly expanding universe, neglecting the effects of the peculiar
velocity. Since $B^{*2}/(8\pi p^{*})<<1$, to an excellent approximation, one
can take the motions induced by the magnetic field to be almost
incompressible, with $p^{*}+B^{*2}/(8\pi )\approx p^{*}={\rm constant}$ and
drop the pressure gradient term in the reduced Euler equation (\ref{euler}).
In this limit we have filtered out ''fast'' compressible motions and we can
look for solutions with ${\bf \nabla }.{\bf v}=0$. Equation (\ref{energ})
then gives, in the ideal and non-viscous limit, $\partial \rho ^{*}/\partial
\tau =0$. Also, the Euler equation reduces to 
\begin{equation}
{\frac{\partial {\bf v}}{\partial \tau }}+({\bf v}.{\bf \nabla }){\bf v}={%
\frac{({\bf B}^{*}.{\bf \nabla }){\bf B}^{*}}{4\pi (\rho ^{*}+p^{*})}};
\label{eulersp}
\end{equation}
while the induction equation is 
\begin{equation}
{\frac{\partial {\bf B}^{*}}{\partial \tau }}+({\bf v}.{\bf \nabla }){\bf B}%
^{*}=({\bf B}^{*}.{\bf \nabla }){\bf v}.  \label{inductsp}
\end{equation}
These equations have a wide class of exact, stable solutions of the form 
\cite{parker79} 
\begin{equation}
{\bf v}=\pm {\frac{{\bf B}^{*}}{\sqrt{4\pi (\rho ^{*}+p^{*})}}},\quad {\frac{%
\partial {\bf v}}{\partial \tau }}=0,\quad {\bf \nabla }.{\bf v}=0
\end{equation}
In these solutions the nonlinear terms in Eq. (\ref{eulersp}) and (\ref
{inductsp}) exactly cancel. Any complicated tangled field pattern is
possible, if accompanied by a velocity along the magnetic field at the local
Alfv\'en speed.

A particular case of this solution is one where the magnetic field is split
up as ${\bf B}^{*}={\bf B}_0^{*}+{\bf b}^{*}$, into a uniform and constant
component, ${\bf B}_0^{*},$ and a tangled component ${\bf b}^{*}({\bf x}%
,\tau )$. Fix the co-ordinates such that ${\bf B}_0^{*}$ lies along the $z$%
-axis, that is ${\bf B}_0^{*}=B_0^{*}\hat {{\bf z}}$, where $\hat {{\bf z}}$
is the unit vector along $z$. Now choose ${\bf v}={\bf b}^{*}/(4\pi (\rho
^{*}+p^{*}))^{1/2}$; the nonlinear terms in the Euler and induction
equations cancel out and we have 
\begin{equation}
{\frac{\partial {\bf b}^{*}}{\partial \tau }}-V_A{\frac{\partial {\bf b}^{*}%
}{\partial z}}=0,\quad {\frac{\partial {\bf v}}{\partial \tau }}-V_A{\frac{%
\partial {\bf v}}{\partial z}}=0,  \label{alfsp}
\end{equation}
where we have defined the Alfv\'en velocity by 
\begin{equation}
V_A={\frac{B_0^{*}}{(4\pi (\rho ^{*}+p^{*}))^{1/2}}}={\frac B{(4\pi (\rho
+p))^{1/2}}}\approx 3.8\times 10^{-4}B_{-9}.  \label{alfvel}
\end{equation}
Here, for the numerical estimate, we have taken the dominant contribution to
the energy density $\rho =\rho _\gamma $, the photon energy density, as
would be appropriate in the radiation-dominated era, after the epoch of $%
e^{+}e^{-}$ annihilation. The general solution of Eq.(\ref{alfsp}) is
therefore a nonlinear Alfv\'en wave travelling antiparallel to ${\bf B}_0^{*}
$, with ${\bf v}={\bf b}^{*}/(4\pi (\rho ^{*}+p^{*}))^{1/2}={\bf F}%
(x,y,z+V_A\tau )$ with an arbitrary function ${\bf F}$. One can also have
another class of solutions with ${\bf v}=-{\bf b}^{*}/(4\pi (\rho
^{*}+p^{*}))^{1/2}$, where the wave travels in the same direction as ${\bf B}%
_0^{*}$. Both these solutions are stable and they exist as long as the two
sets of waves do not overlap in space \cite{parker79}.

It is not possible to generalise these nonlinear solutions to the diffusive
viscous regime, in the above form, for arbitrary viscosity and conductivity
coefficients. However, there exists a special case (see Landau and Lifshitz 
\cite{ll}), with the velocity and tangled magnetic field having arbitrary
strengths, but aligned perpendicular to ${\bf B}_0^{*}$ and depending only
on $z$, where such a generalisation is possible. We now look at this case in
more detail.

\subsection{Nonlinear Alfv\'en waves in the viscous regime}

We begin by reinstating the viscous term in the Euler equation (\ref{euler}%
). As before, assume that the magnetic field can be written as ${\bf B}^{*}=%
{\bf B}_0^{*}+{\bf b}^{*}$, with a uniform ${\bf B}_0^{*}$. We assume ${\bf b%
}^{*}$ is perpendicular to ${\bf B}_0^{*}$, but do not put any restriction
on the strength of ${\bf b}^{*}$ so that it need {\it not be a small}
perturbation of ${\bf B}_0^{*}$. We also take the peculiar velocity ${\bf v}$
to lie perpendicular to ${\bf B}_0^{*}$ and assume that all the variables
depend only on $z$ and $\tau $. In this case, the velocity perturbation
automatically satisfies ${\bf \nabla }.{\bf v}=0$. Further, the ratio of the
magnetic energy density to the fluid energy density, $B^2/(8\pi \rho )\sim
10^{-6}B_{-9}^2<<1$. So even when there is strong damping of motions induced
by the field, and a significant fraction of the field energy density goes
into heat, $\rho $ will be perturbed negligibly. It is an excellent
approximation to neglect the viscous term in Eq. (\ref{energ}). Then, in the
ideal limit, Eq. (\ref{energ}) implies $(\partial \rho ^{*}/\partial \tau
)=0 $. The non-linear terms in the Euler and induction equations are
individually zero because there is no variation of ${\bf b}^{*}$ and ${\bf v}
$ along the fields. These equations then reduce to 
\begin{equation}
{\frac{\partial {\bf v}}{\partial \tau }}=-\left( {\frac 1{\rho ^{*}+p^{*}}}%
\right) {\bf \nabla }\left[ p^{*}+B^{*2}/(8\pi )\right] +\left( {\frac{%
B_0^{*}}{4\pi (\rho ^{*}+p^{*})}}\right) {\frac{\partial {\bf b}^{*}}{%
\partial z}}+{\frac{\eta ^{*}}{(\rho ^{*}+p^{*})}}\nabla ^2{\bf v}
\label{eulerred}
\end{equation}
\begin{equation}
{\frac{\partial {\bf b}^{*}}{\partial \tau }}=B_0^{*}{\frac{\partial {\bf v}%
}{\partial z}}  \label{inductred}
\end{equation}
Note that the LHS of Eq. (\ref{eulerred}) has zero divergence. The RHS will
also have zero divergence only if the total pressure $p*+B^{*2}/(8\pi )$ is
uniform in space. As mentioned above, this is likely to be a good
approximation for this mode since the radiation pressure in the early
universe is typically much larger than the magnetic pressure. One can
therefore drop the pressure gradient term in the reduced Euler equation (\ref
{eulerred}). Writing ${\bf b}^{*}=b_0(\tau ,z){\bf n}$ and ${\bf v}=v_0(\tau
,z){\bf n}$, eliminating $v_0$ from Eqns. (\ref{eulerred}) and (\ref
{inductred}), gives a damped wave equation for $b_0(\tau ,z)$, 
\begin{equation}
{\frac{\partial ^2b_0}{\partial \tau ^2}}-{\frac{\eta ^{*}(\tau )}{(\rho
^{*}+p^{*})}}{\frac \partial {\partial z^2}}\left( {\frac{\partial b_0}{%
\partial \tau }}\right) -V_A^2{\frac{\partial ^2b_0}{\partial z^2}}=0,
\label{almfin}
\end{equation}
where we have defined the Alfv\'en velocity, $V_A,$ as before. This linear
equation generalises the nonlinear Alfv\'en mode to the viscous regime. It
can easily be solved by taking a spatial Fourier transform. For any mode $%
b_0(\tau ,z)=f(\tau )e^{ikz}$. we have 
\begin{equation}
{\ddot f}+{\frac{\eta ^{*}(\tau )k^2}{(\rho ^{*}+p^{*})}}{\dot f}+k^2V_A^2f=0
\label{oscill}
\end{equation}
which is the equation for a damped harmonic oscillator.

The behaviour of solutions to the damped oscillator equation depends on the
relative strengths of the driving and damping terms. Suppose we define 
\begin{equation}
\omega _0=kV_A;\quad D={\frac{\eta ^{*}(\tau )k^2}{(\rho ^{*}+p^{*})}\propto 
}\tau ^2.  \label{domdef}
\end{equation}
If $\omega _0>>D$, then we will have damped oscillatory motion. In the other
extreme limit of $D>>\omega _0$, the motion becomes overdamped. While one
solution of the second-order differential equation suffers strong damping,
the other independent solution is negligibly damped . The physical reason
for this is that, under strong friction, any oscillator displaced from
equilibrium and released from rest has only to acquire a small ''terminal''
velocity, so that friction balances the driving force. An oscillator
starting from this ''phase'' of oscillation, will then almost freeze, and
the associated energy in the oscillator will decrease negligibly. On the
other hand, an oscillator mode with a large initial velocity will be
significantly damped by the strong friction. Therefore, it is important to
consider the ratio $D/\omega _0$ to determine which limit applies for the
nonlinear Alfv\'en modes.

We focus primarily on damping by photon viscosity. This is the most
important source of viscosity, after $e^{+}e^{-}$ annihilation. Also, it is
the dissipative process with the potential to damp the largest scales.
(Smaller-scale damping by neutrinos is briefly discussed in Section VII and
Appendix C). The radiative viscosity coefficient is given by Eq. (\ref
{etabef}), with $g=g_\gamma =2,$ and the photon mean-free-path is 
\begin{equation}
l_\gamma (\tau )={\frac 1{\sigma _Tn_e(\tau )}}\approx 9.5\times
10^{21}cm\left( {\frac T{0.25eV}}\right) ^{-3}\left( {\frac{\Omega _bh^2}{%
0.0125}}\right) ^{-1}x_e^{-1}.  \label{lgamnum}
\end{equation}
Here, $\sigma _T$ is the Thomson cross-section for electron-photon
scattering, $n_e$ is the electron number density, $x_e$ the ionisation
fraction, and $\Omega _b$ is the baryon density of the universe $\rho _b$,
in units of the closure density. (For later convenience we define the
quantity $f_b\equiv (\Omega _b/0.0125h^{-2})$, which measures the baryon
density, in terms of the preferred value given by Walker {\it et al.}\cite
{walker}. This used to be the canonical value determined from
nucleosynthesis constraints, although at present there is some debate on
this issue \cite{hata}). In the early universe, the energy density of the
baryon-photon fluid is dominated by the photon energy, density $\rho _\gamma 
$; so, using $p^{*}=\rho ^{*}/3$, $\rho =\rho ^{*}/a^4$, $\eta ^{*}=a^3\eta $%
, we have for the damping coefficient 
\begin{equation}
D={\frac{\eta ^{*}k^2}{(\rho ^{*}+p^{*})}}=a^3{\frac 4{15}}{\frac{\rho
_\gamma (\tau )l_\gamma (\tau )k^2}{(4\rho _\gamma (\tau )a^4/3)}}={\frac 15}%
k^2l_\gamma (\tau _0)a^2(\tau ),  \label{dam}
\end{equation}
where we have used the fact that $l_\gamma (\tau )\propto n_e^{-1}(\tau
)\propto a^3(\tau )$ and, as before, $\tau _0$ is the conformal time today.
Hence, the damping-to-driving ratio is 
\begin{eqnarray}
{\frac D{\omega _0}} &=&{\frac{\eta ^{*}k^2}{kV_A(\rho ^{*}+p^{*})}}={\frac 1%
5}{\frac{k_p(\tau )l_\gamma (\tau )}{V_A}}  \nonumber \\
\ &\approx &526.3{\frac{k_p(\tau )l_\gamma (\tau )}{B_{-9}}}.  \label{bbyom}
\end{eqnarray}
We have defined the proper wavenumber $k_p(\tau )=(k/a(\tau ))$, of a
Fourier component and substituted for the Alfv\'en velocity in terms of the
field strength using $V_A\approx 3.8\times 10^{-4}B_{-9}$. For the diffusion
approximation to be valid, we require $k_pl_\gamma <1$; that is, we must
consider only wavelengths larger than the mean-free-path. Nevertheless, one
expects a large range of wavelengths for which modes will fall in the
overdamped regime.

In order to consider the evolution of modes of different wavelengths, one
has first to look at the quantitative solution of Eq. (\ref{oscill}) . For
this, substituting 
\begin{equation}
f(\tau )=\exp \left( -\int {\frac{D(\tau )}2}d\tau \right) W(\tau )
\label{subst}
\end{equation}
into Eq. (\ref{oscill}), the evolution of $W$ is given by 
\begin{equation}
{\ddot W}+p(\tau )W=0\quad {\rm with}\quad p(\tau )=\omega _0^2-{\frac{{\dot 
D}}2}-({\frac D2})^2 .  \label{Weq}
\end{equation}
When $\omega _0>>D$, we have $p\approx \omega _0^2$ in Eq. (\ref{Weq}) and
then the solution is $W=\exp (\pm i\omega _0\tau )$. Therefore, in the
oscillatory limit we have 
\begin{equation}
b_0(\tau ,z)=exp\left( -\int {\frac{D(\tau )}2}d\tau \right) e^{\pm i\omega
_0\tau +ikz}, \quad \omega _0>>D.  \label{oslim}
\end{equation}
In the opposite limit, $D>>\omega _0$, we have to solve the oscillator with
a time-dependent friction coefficient. One can obtain an approximate WKBJ
solution, 
\begin{equation}
W(\tau )={\frac 1{(-p)^{1/4}}}\exp \left[ \pm \int (-p)^{1/2}(\tau )d\tau
\right]  \label{Wsol}
\end{equation}
This solution is valid as long as $p(\tau )$ does not vary too rapidly. In
the overdamped regime, this condition can be shown to be equivalent to
neglecting ${\dot D}$ compared to $D^2$. In the limit ${\dot D}<< D^2$, the
two solutions are given by 
\begin{equation}
f(\tau )=A_0{\frac 1{D^{1/2}}}\exp \left( -\int^\tau D(\tau ^{\prime })d\tau
^{\prime }\right) ;\quad f(\tau ) =B_0{\frac 1{D^{1/2}}}\exp \left(
-\int^\tau {\frac{\omega _0^2}{D(\tau ^{\prime })}}d\tau ^{\prime }\right)
\label{wkbover}
\end{equation}
As advertised, in the overdamped limit, one solution (the $A_0$ - mode) is
strongly damped while the other solution (with $B_0\ne 0$) is weakly damped.

For damping by photon viscosity, we have $D\propto a^2\propto (\tau /\tau
_0)^2$ in the radiation-dominated epoch. It is more useful to consider an
alternate treatment to that of the WKBJ solution, for the rapidly varying,
strongly overdamped regime. One notes that, as the damping increases with
time to $D>>\omega _0$, $\dot f$ will tend to adjust itself so that the
acceleration vanishes, so $\ddot f\approx 0$. For example, consider
initially the case $f>0,\dot f<0$ and $D\dot f>\omega _0^2f$. Then $\ddot f%
>0 $ and so the magnitude of $\dot f$ decreases (while remaining negative)
until we have $D\dot f=-\omega _0^2f$, when the acceleration vanishes. On
the other hand, if $D\dot f<\omega _0^2f$, then $\ddot f<0$ and the
magnitude of $\dot f$ increases until we have $\ddot f=0$. Subsequently as $%
D $ keeps increasing, $\dot f$ can continue to adjust itself to maintain
zero $\ddot f$. One can argue similarly for all other cases. Therefore, it
seems plausible to consider an alternate approximation for the overdamped
case, whereby, after the time when $\ddot f$ first vanishes, $f$ satisfies
the equation 
\begin{equation}
D\dot f+\omega _0^2f=0;\quad D>>\omega _0.  \label{termapp}
\end{equation}
We will refer to this approximation as the terminal-velocity approximation.
The solution in the overdamped regime, under the terminal-velocity
approximation, is simply given by 
\begin{equation}
f(\tau )=f(\tau _T)\exp \left( -\int_{\tau _T}^\tau {\frac{\omega _0^2}{%
D(\tau ^{\prime })}}d\tau ^{\prime }\right) .  \label{tersol}
\end{equation}
Here, $\tau _T$ is the conformal time when the mode reaches the
terminal-velocity regime, or when the acceleration, $\ddot f,$ first
vanishes. As we explained earlier, for an oscillator with an initial phase
such that $\dot f$ is already large, this implies strong damping by the time
the terminal-velocity regime is reached. On the other hand, for an
oscillator which starts from rest, $\dot f$ will have to increase negligibly
for the $D>>\omega _0$ regime, so that $\ddot f$ vanishes and Eq. (\ref
{tersol}) applies.

We now move from the study of the non-linear Alfv\'en modes to consider the
damping of all the different MHD modes in their linearised limit. This has
already been studied by JKO\cite{JKO96}. However, we shall do this using the
formalism developed here to bring out the links with the non-linear
situation. We will return to the further evolution of the non-linear
Alfv\'en mode in Section VI.

\section{ Damping of linearised compressible MHD waves}

Let us begin with the linearised MHD equations describing small
perturbations to density $\rho _1^{*}=\rho ^{*}-\rho _0^{*}$, pressure $%
p_1^{*}=p^{*}-p_0^{*},$ and magnetic field ${\bf b}^{*}={\bf B}^{*}-{\bf B}%
_0^{*}$. We have 
\begin{equation}
{\frac{\partial \rho _1^{*}}{\partial \tau }}+{\bf \nabla }.[(\rho
_0^{*}+p_0^{*}){\bf v}]=0.  \label{energl}
\end{equation}
\begin{equation}
{\frac \partial {\partial \tau }}\left[ (\rho _0^{*}+p_0^{*}){\bf v}\right]
=-{\bf \nabla }p_1^{*}+{\frac{[{\bf \nabla }\times {\bf b}^{*}]\times {\bf B}%
_0^{*}}{4\pi }}+\eta ^{*}\left[ \nabla ^2{\bf v}+{\frac 13}{\bf \nabla }(%
{\bf \nabla }.{\bf v})\right] .  \label{eulerl}
\end{equation}
\begin{equation}
{\frac{\partial {\bf b}^{*}}{\partial \tau }}={\bf \nabla }\times \left[ 
{\bf v}\times {\bf B}_0^{*}\right]  \label{inductl}
\end{equation}
\begin{equation}
{\bf \nabla }.{\bf b}^{*}=0,\quad p_0^{*}={\frac{\rho _0^{*}}3},\quad
p_1^{*}={\frac{\rho _1^{*}}3.}  \label{otherl}
\end{equation}
Suppose the perturbation is described in terms of the perturbed comoving
position $\delta {\bf x}={\bf \xi }({\bf x},\tau )$. The perturbed velocity $%
{\bf v}=(\partial {\bf \xi }/\partial \tau )\equiv {\dot {{\bf \xi }}}$. An
integration of the perturbed continuity equation (\ref{energl}) and
induction equation (\ref{inductl}) then gives 
\begin{equation}
\rho _1^{*}=-{\frac 43}\rho _0^{*}{\bf \nabla }.{\bf \xi }\;\quad {\bf b}%
^{*}={\bf \nabla }\times \left[ {\bf \xi }\times {\bf B}_0^{*}\right] .
\label{rhobl}
\end{equation}
Substituting Eq. (\ref{rhobl}) into the perturbed Euler equation, we get 
\begin{equation}
{\ddot {{\bf \xi }}}={\frac 13}{\bf \nabla }({\bf \nabla }.{\bf \xi }%
)+\left[ {\bf \nabla }\times ({\bf \nabla }\times [{\bf \xi }\times {\bf v}%
_A])\right] \times {\bf v}_A+{\frac{3\eta ^{*}}{4\rho _0^{*}}}\left[ \nabla
^2{\dot {{\bf \xi }}}+{\frac 13}{\bf \nabla }({\bf \nabla }.{\dot {{\bf \xi }%
}})\right] ,  \label{linper}
\end{equation}
where we have defined ${\bf v}_A=V_A\hat {{\bf z}}$. This linear equation
describes the evolution and damping of linearised MHD modes in the expanding
universe. One can look for plane wave solutions of the form ${\bf \xi }={\bf %
\psi }(\tau )\exp (i{\bf k}.{\bf x})$. This leads to a replacement of ${\bf %
\nabla }$ by $i{\bf k}$ and leads to an evolution equation for the amplitude 
${\bf \psi }$: 
\begin{equation}
{\ddot {{\bf \psi }}}=-{\bf k}\left[ ({\bf k}.{\bf \psi })[{\frac 13}%
+V_A^2]-V_A^2k_z\psi _z\right] -V_A^2k_z^2{\bf \psi }-k_z({\bf k}.{\bf \psi }%
)V_A^2\hat {{\bf z}}-{\frac{3\eta ^{*}}{4\rho _0^{*}}}\left[ k^2{\dot {{\bf %
\psi }}}+{\frac 13}{\bf k}({\bf k}.{\dot {{\bf \psi }}})\right] .
\label{linperk}
\end{equation}
Here a subscript ''z'' denotes the $z$-component of the relevant quantity.
We can now look at various types of solution to the above equation.

First, consider the incompressible mode, with ${\bf \nabla }.{\bf \xi }=0$,
or ${\bf k}.{\bf \psi }=0$. In this case taking the dot product of Eq. (\ref
{linperk}) with ${\bf k}$, we also have $\psi _z=0$ (provided $k_z\equiv
k\cos {\theta }\ne 0$). For this mode, Eq. (\ref{linperk}) reduces to 
\begin{equation}
{\ddot {{\bf \psi }}}+{\frac{3\eta ^{*}}{4\rho _0^{*}}}k^2{\dot {{\bf \psi }}%
}+V_A^2\cos ^2{\theta }k^2{\bf \psi }=0.  \label{linalf}
\end{equation}
This is, as expected, is almost exactly the evolution equation for the
Alfv\'en mode encountered and discussed in detail in the last section. The
only difference is that $V_A$ is replaced by $V_A\cos {\theta }$,
generalising the Alfv\'en mode propagation to be in a general direction
inclined at an angle $\theta $ to the zero-order magnetic field. (Although
this generalisation is at the cost of introducing the linear approximation).

The evolution of the compressible modes, can be derived by taking the dot
product of Eq. (\ref{linperk}) with ${\bf k}$ and $\hat {{\bf z}}$. Defining 
$A={\bf k}.{\bf \psi }/k$, we have 
\begin{equation}
{\ddot A}+[{\frac 13}+V_A^2]k^2A-V_A^2k^2\cos ^2{\theta }\psi _z+{\frac{\eta
^{*}}{\rho _0^{*}}}k^2{\dot A}=0,  \label{linA}
\end{equation}
\begin{equation}
{\ddot \psi _z}+{\frac 13}k^2\cos {\theta }A+{\frac{3\eta ^{*}}{4\rho _0^{*}}%
}k^2{\dot \psi _z}+{\frac{\eta ^{*}}{4\rho _0^{*}}}k^2\cos {\theta }{\dot A}%
=0.  \label{linpsiz}
\end{equation}
Consider first the undamped limit with $\eta ^{*}=0$. In this case, looking
for modes with $A\propto \psi _z\propto e^{i\omega \tau }$, we can easily
derive the dispersion relations 
\begin{equation}
{\frac{\omega ^2}{k^2}}={\frac 12}\left[ c_s^2+V_A^2\right] \pm {\frac 12}%
\left[ (c_s^2+V_A^2)^2-4c_s^2V_A^2\cos ^2{\theta }\right] ^{1/2}.
\label{displ}
\end{equation}
Here, we have defined the sound speed in the relativistic limit $c_s=1/\sqrt{%
3}$. The plus sign in the above equation corresponds to the fast MHD mode
while the negative sign corresponds to the slow mode. In the limit $V_A<<c_s$%
, which is generally applicable to our early universe context, the
dispersion relation for the fast mode becomes $\omega /k\approx c_s$, while
that of the slow mode becomes $\omega /k\approx V_A\cos {\theta }$.

The general solution of Eq. (\ref{linA}) and (\ref{linpsiz}) when the
damping terms are reinstated is quite complicated to analyze analytically,
since it involves a fourth-order differential equation with time-dependent
coefficients and the dispersion relation is a fourth-order polynomial.
However, we can look at some simple special cases which illustrate the
general behaviour.

First, consider the case where ${\bf k}$ is parallel to ${\bf B}_0^{*}$;
then $\cos {\theta }=k_z/k=1$, $A=\psi _z$, and the equation for $A$ reduces
to 
\begin{equation}
{\ddot A}+{\frac{\eta ^{*}}{\rho _0^{*}}}k^2{\dot A}+c_s^2k^2A=0.
\label{linsA}
\end{equation}
This describes a damped sound wave, well studied in the literature in
connection with the Silk damping of acoustic baryon-photon fluctuations.
Therefore we only look at it briefly, to estimate the Silk damping scale.
Because the sound-wave oscillation frequency is such that $\omega
_s=kc_s>>kV_A$, these modes {\it do not} become overdamped in general, and
the damped oscillatory solutions of Section IV can be used to describe their
evolution. Specifically, we have 
\begin{equation}
A(\tau )=exp\left( -\int {\frac{D_s(\tau )}2}d\tau \right) e^{\pm i\omega
_s\tau },\quad \omega _s>>D_s  \label{oslims}
\end{equation}
where $D_s=k^2(\eta ^{*}/\rho _0^{*})$. These modes get damped by a factor 
\begin{equation}
exp\left( -\int {\frac{D_s(\tau )}2}d\tau \right) =\exp \left[ -{\frac{k^2}{%
k_D^2}}\right] ;\quad {\rm where}\quad k_D^{-2}={\frac 2{15}}\int {\frac{%
l_\gamma dt}{a^2(t)}}.  \label{sldamp}
\end{equation}
This agrees quite well with the Silk damping of sound waves in the radiation
era, derived in more detailed treatments (cf. \cite{peeb}, \cite{kaiser}),
in the appropriate limit. In fact, the more detailed derivation of Silk
damping, using a Boltzmann treatment gives a damping factor $\exp
[-(k^2/k_{D,bol}^2)]$, where 
\begin{equation}
k_{D,bol}^{-2}={\frac 2{15}}\int {\frac{l_\gamma dt}{a^2(t)}}\left[ {\frac{%
(1+R)f_2^{-1}+5R^2/4}{(1+R)^2}}\right] .  \label{sldampus}
\end{equation}
Here, $R=3\rho _b/4\rho _\gamma $ and $f_2=3/4$ if the effects of
polarisation are included and $f_2=1$ otherwise. Notice that, in the limit $%
R<<1$ applicable to epochs where the radiation density dominates baryon
density, and $f_2=1$, the damping scales exactly match, with $k_D=k_{D,bol}$%
. The effects of non-zero $R$ constitute at most about a $20-25\%$
correction to the damping scale we derive. In the radiation-dominated epoch
one has $k_D^{-1}=(4/45)^{1/2}L_S(t)/a(t)\sim 0.3L_S(t)/a(t)$, where $%
L_S(t)=(l_\gamma t)^{1/2}$ is the Silk scale. The largest scales which
suffer appreciable damping are the modes with wavelengths ($2\pi k_D^{-1})$,
of order $L_S$ \cite{silk}.

In the other extreme case, when ${\bf k}$ is perpendicular to ${\bf B}_0^{*}$%
, $\cos {\theta }=k_z/k=0$, and the equation for $A$ reduces to 
\begin{equation}
{\ddot A}+{\frac{\eta ^{*}}{\rho _0^{*}}}k^2{\dot A}+[{\frac 13}%
+V_A^2]k^2A=0.  \label{linpA}
\end{equation}
This describes a damped fast-magnetosonic wave. The real part of the
oscillation frequency is $\omega _R=k(c_s^2+V_A^2)^{1/2}>>kV_A$, since in
general $V_A<<c_s$. One can see that the damping of these modes is very
similar to that of the sound waves (and in fact is exactly the same when we
neglect $V_A$ compared to $c_s$). Again, we expect modes with wavelengths
less than the Silk scale to be significantly damped.

Now we turn to the damping of modes with arbitrary direction {\bf k}. In the
undamped case, with $\eta ^{*}=0$, and when $V_A<<c_s$, we have seen that
the fast mode has the same oscillation frequency as the sound wave and the
slow mode as the Alfve\'n wave. This suggests an approximation to capture
the damped counterparts of these modes in the limit of weak magnetic field.
Let us write the time variation of $A$ and $\psi _z$ as $A(\tau )\propto
\psi _z\propto \exp {(i\int \omega d\tau )}$. Suppose $\omega $ is dominated
by its real part, and this is of order the undamped frequency of
oscillation. We shall later check the consistency of this assumption.
Consider first the damped counterpart of the slow mode, in the limit $%
V_A^2<<c_s^2=1/3$. For this mode, the ratio of the first two terms in Eq. (%
\ref{linA}), is ${\ddot A}/(c_s^2k^2A)\sim V_A^2/c_s^2<<1$. Also, the ratio
of the last term in (\ref{linA}) to the second is $\sim D{\dot A}%
/(c_s^2k^2A)\sim V_Ak_pl_\gamma /5<<1$. So, these two terms can be neglected
when compared to the second term in Eq. (\ref{linA}), so $c_s^2k^2A\approx
V_A^2k^2\cos {\theta }\psi _z$. For the same reason, one can neglect the
last term in Eq. (\ref{linpsiz}) compared to the second term in this
equation. Substituting $c_s^2k^2A\approx V_A^2k^2\cos {\theta }\psi _z$ in
Eq. (\ref{linpsiz}), we then have for the damped counterpart of the slow
mode, 
\begin{equation}
{\ddot \psi _z}+{\frac{3\eta ^{*}}{4\rho _0^{*}}}k^2{\dot \psi _z}%
+V_A^2k^2\cos ^2{\theta }\psi _z=0.  \label{linpsidz}
\end{equation}
We see that this is exactly the same equation as that obtained for the
damped Alfv\'en mode, analysed in the last section. Therefore, the slow
modes will also be overdamped, and have one solution with negligible damping
rate. Our original assumption that $\omega $ is dominated by its real part
is valid for this solution, showing the self-consistency of our assumptions.
Also, for this mode we have $A\approx (V_A^2/c_s^2)\cos {\theta }\psi
_z<<\psi _z$, for a general $\theta $; that is, the mode is almost
incompressible. (The strongly damped mode has to be analysed differently).

Now consider the fast mode in a similar fashion. We have already derived the
exact evolution for the special case when $\cos {\theta }=0$. Suppose $\cos {%
\theta }\ne 0$; assume, as before, that $\omega $ is dominated by its real
part, and this is of order the undamped frequency of oscillation. Then,
substituting for the time dependence of $A$ and $\psi _z$, and taking the
real part of Eq. (\ref{linpsiz}), we have $\psi _z\sim A\cos {\theta }$. So
the ratio of the $\psi _z$ term in Eq. (\ref{linA}) compared to the second
term is $\sim (V_A^2/c_s^2)\cos ^2{\theta }<<1$. Neglecting the $\psi _z$
term, and neglecting $V_A^2$ compared to $c_s^2$ in Eq. (\ref{linA}),
results once again in the same equation, Eq. (\ref{linsA}), for the fast
wave as was found for the damped sound waves.

In summary, the above analysis for linear MHD waves, in the limit of weak
fields with $V_A << c_s$, shows that the fast magnetosonic waves generally
damp like sound waves, while there is one mode of the slow magnetosonic wave
which behaves exactly as the Alfve\'n mode and gets overdamped. This also
agrees with the conclusions reached by JKO.

\section{The free-streaming regime}

As the universe expands, the mean-free-path of the photon increases as $a^3$%
, while the proper length of any perturbed region increases as $a$. So the
photon mean-free-path can eventually become larger than the proper
wavelength of a given mode. When this happens for any given mode, we will
say that the mode has entered the free-streaming regime. Modes with
progressively larger wavelengths enter the free-streaming regime up to a
proper wavelength $\sim l_\gamma (T_d)\sim 10^{22}cm$ (see Eq. (\ref{lgamnum}%
) ), or a comoving wavelength of $\sim 3$Mpc, at the epoch of decoupling.
After (re)combination of electrons and nuclei into atoms, $l_\gamma $
increases to a value larger than the present Hubble radius, and all modes
enter the free-streaming regime. (We will consider the pre- and
post-recombination epochs separately below.)

When photons start to free stream on a given scale of perturbation, the
tight-coupling diffusion approximation no longer provides a valid
description of the evolution of the perturbed photon-baryon fluid on that
scale. One has to integrate the Boltzmann equation for the photons together
with the MHD equations for the baryon-magnetic field system. A simpler
approximate method of examining the evolution of such modes in the linear
regime is to treat the radiation as isotropic and homogeneous, and only
consider its frictional damping force on the fluid. (The radiative flux
could have also contributed to the force on the baryons; however, for modes
with wavelengths smaller than $l_\gamma $, this flux is negligible since the
associated compressible motions have suffered strong Silk damping at earlier
epochs; when the wavelength was larger than $l_\gamma $). The drag force on
the baryon fluid per unit volume due to the radiation energy density $\rho
_\gamma $, is given by 
\begin{equation}
{\bf F}_D=-{\frac 43}n_e\sigma _T\rho _\gamma {\bf v.}  \label{dragfor}
\end{equation}

Since, typically, less than one electron-photon scattering occurs within a
wavelength, the pressure and inertia contributed by the radiation can be
neglected when considering the evolution of such modes. The Euler equation
for the baryonic component then becomes 
\begin{equation}
{\frac{\partial {\bf v}}{\partial t}}+H(t){\bf v}+{\bf v}.{\bf \nabla }{\bf v%
}=-{\frac 1{a\rho _b}}{\bf \nabla }p_b+{\frac 1{\rho _b}}{\bf J}\times {\bf B%
}-{\frac 1a}{\bf \nabla }\phi -{\frac{4\rho _\gamma }{3\rho _b}}n_e\sigma _T%
{\bf v.}  \label{eulerin}
\end{equation}
Here, $\rho _b$ is the baryon density, $p_b$ the fluid pressure, and $%
H(t)=(da/dt)/a$ is the Hubble parameter. We have also included the
gravitational force, $(1/a){\bf \nabla }\phi $, due to any perturbation in
the density. Note that we have written this equation in the unstarred
conformal frame, (with the magnetic field defined in the ''Lab'' frame; see
Appendix A), since conformal transformation to flat spacetime is no longer a
useful tool in the matter-dominated era. We have also transformed the time
co-ordinate, from conformal time, back to ''proper time'' $dt=ad\tau $.

It should be pointed out that the dramatic drop in the pressure, by a factor
of order the very small baryon to photon ratio $\sim 10^{-9}$, when a mode
enters the free-streaming regime, has important consequences. First, in the
absence of radiation pressure, the effect of magnetic pressure (if it
greatly exceeds the fluid pressure) is to convert what was initially an
incompressible Alfve\'n mode into a compressible mode (see below). Second,
the effective baryonic Jeans mass decreases dramatically and compressible
modes can become gravitationally unstable. Thus, we have to retain the
gravitational force term in the above equation. The magnetic pressure will
also play a dominant role, providing pressure support against gravity on
sufficiently small scales.

The evolution of modes which enter the free-streaming regime depends on the
strength of the magnetic fields, in particular whether the magnetic pressure 
$p_B$ is greater or smaller than the fluid pressure, $p_b$. For the magnetic
pressure, we have 
\begin{equation}
p_B={\frac{B^2}{8\pi }}(1+z)^4\approx 4\times 10^{-8}B_{-9}^2\left( {\frac{%
1+z}{10^3}}\right) ^4dyn/cm^2.  \label{magp}
\end{equation}
While the fluid pressure is given by 
\begin{equation}
p_b=2n_ekT\approx 1.1\times 10^{-10}\left( {\frac{1+z}{10^3}}\right) ^4f_b%
\text{ }dyn/cm^2,  \label{fluidp}
\end{equation}
where we have assumed that the fluid temperature is locked to the radiation
temperature, and that the gas is an electron-proton gas. By taking the ratio
of the two pressures, one can see that magnetic pressure dominates the fluid
pressure, (i.e. $p_B>>p_b$ for $B>>B_{crit}\sim 5\times 10^{-11}$ Gauss).
For magnetic fields smaller than $B_{crit}$, the fluid pressure dominates.

Consider first the case where the field $B$ is much smaller than $B_{crit}$.
In this case the motions can be assumed incompressible. The Alfv\'en modes
which enter the free-streaming regime, remain Alfv\'enic. Following the
ideas of Section IV, we look again at non-linear Alfv\'en modes with ${\bf B}%
=({\bf B}_0+{\bf b})/a^2$, where ${\bf B}_0=B_0\hat {{\bf z}}$, with $B_0=%
{\rm constant}$, ${\bf b}={\bf n}\bar b_0(z,t)$ and ${\bf v}={\bf n}\bar v%
_0(z,t)$, with ${\bf n}$ perpendicular to $\hat {{\bf z}}$. Recall that $|%
{\bf b}|$ is {\it not necessarily small} compared to $|{\bf B}_0|$. We
assume $\rho _b$ to be uniform (but not independent of $t$), use the Euler
equation (\ref{eulerin}) and the induction equation (\ref{ind}), change to
conformal time $\tau $, and look for solutions in the form $\bar b_0(z,t)={%
\bar f}(t)e^{ikz}$, following the same procedure as in Section IV. (For the
rotational Alfv\'en-type mode, the gradient terms in (\ref{eulerin}) do not
contribute). We obtain an equation for the evolution of ${\bar f}(\tau )$ 
\begin{equation}
{\frac{d^2{\bar f}}{d\tau ^2}}+\left[ aH+\bar D\right] {\frac{d{\bar f}}{%
d\tau }}+\bar \omega _0^2{\bar f}=0,  \label{barf}
\end{equation}
where 
\begin{equation}
\bar \omega _0=kV_A({\frac{4\rho _\gamma }{3\rho _b}})^{1/2};\quad \bar D%
=n_e\sigma _Ta({\frac{4\rho _\gamma }{3\rho _b}}).  \label{bardomdef}
\end{equation}
Here, $V_A^2=3B_0^2/(16\pi \rho _\gamma )={\rm constant}$, is the same as
defined in terms of the starred variables in Section IV.

The evolution of this non-linear Alfv\'en mode depends once again on the
relative strengths of the damping and driving terms. The ratio of the
viscous damping to expansion damping is given by 
\begin{equation}
{\frac{\bar D}{aH}}={\frac{(4\rho _\gamma /3\rho _b)n_e\sigma _Ta}{aH}}={%
\frac{4\rho _\gamma }{3\rho _b}}{\frac{D_H}{l_\gamma }}>>1  \label{scatexp}
\end{equation}
since the Hubble radius $D_H\equiv H^{-1}>>l_\gamma $, so one can neglect
the damping due to Hubble expansion. Also, the ratio of the viscous damping
to the driving terms in the oscillator equation (\ref{barf}) is 
\begin{equation}
{\frac{\bar D}{\bar \omega _0}}={\frac{(4\rho _\gamma /3\rho _b)n_e\sigma _Ta%
}{kV_A(4\rho _\gamma /3\rho _b)^{1/2}}}\approx 3.04\times 10^3({\frac{\rho
_\gamma }{\rho _b}})^{1/2}{\frac 1{k_p(t)l_\gamma (t)B_{-9}}}.
\end{equation}
When a given mode enters the free-streaming limit we will have $%
k_p(t)l_\gamma (t)\sim 1$. So, for the field strengths $%
B_{-9}<(B_{crit}/10^{-9}G)<<1$ that we are considering, all the Alfv\'en
modes are strongly overdamped. As the universe expands, the product $%
k_p(t)l_\gamma (t)\propto a^2$ increases, and at late times any given mode
enters the damped oscillatory regime. One can again apply the
terminal-velocity approximation of Section IV in the overdamped regime once $%
d{\bar f}/d\tau $ has adjusted itself to the ''zero acceleration'' solution.
Then, $\bar f$ is given by 
\begin{equation}
{\bar f}(\tau )={\bar C}\exp \left( -\int^\tau {\frac{\bar \omega _0^2(\tau
^{\prime })}{{\bar D}(\tau ^{\prime })}}d\tau ^{\prime }\right) ;\quad \bar D%
>>\bar \omega _0.  \label{terovert}
\end{equation}
At sufficiently late times, such that $\bar D/\bar \omega _0<<1$, the damped
oscillatory solutions are appropriate, so 
\begin{equation}
\bar f=exp\left( -\int {\frac{\bar D(\tau )}2}d\tau \right) e^{\pm i\bar 
\omega _0\tau },\quad \bar \omega _0>>\bar D.  \label{oslimt}
\end{equation}

Now consider the case $B>>B_{crit}$, when magnetic pressure dominates over
the fluid pressure. For $p_B>>p_b$, incompressibility is no longer a good
assumption, and we necessarily will also have gravitationally unstable,
magneto-acoustic modes. In general, the pressure-gradient term (dominated by
the magnetic field) and the magnetic tension are of similar order. The
non-uniform magnetic fields (associated with what were initially
Alfv\'en-type modes), can seed compressible motions and their associated
density fluctuations. These density fluctuations, can grow by gravitational
instability into non-linear structures. Before recombination, there is still
significant drag due to the free-streaming photons. But once atoms form, the
density of free electrons decreases and the photon mean-free-path becomes
larger than the Hubble radius. In the free-streaming regime, the evolution
equations of these gravitationally unstable MHD modes are the same before
and after recombination. We consider a unified analysis of this evolution in
the next section.

\section{ The post-recombination regime}

We assume that the perturbations in density and velocity are small enough so
that non-linear terms in the perturbed density and velocity can be
neglected. In the Euler equation (\ref{eulerin}), one can neglect the
non-linear term, ${\bf v}.{\bf \nabla }{\bf v}$ and take the density $\rho
_b $ to be that of the unperturbed FRW background density. This equation has
to be supplemented by the continuity equation for the perturbed fluid
density, the Poisson equation for the potential, and the induction equation (%
\ref{ind}). These equations are given explicitly in Appendix B. In the
Poisson equation, we take account of the possibility that there may be other
forms of collisionless dark matter by writing 
\begin{equation}
{\bf \nabla }^2\phi =4\pi Ga^2\delta \rho _T=4\pi Ga^2\left[ \rho _b\delta
_b+\rho _c\delta _c\right] .  \label{poison}
\end{equation}
Here, $\delta \rho _T$ is the total perturbed density (due to both fluid
plus dark matter), $\delta _b$ is the fractional perturbation in the fluid,
while $\rho _c$ and $\delta _c$ describe the dark matter density and its
fractional perturbation, respectively. We shall adopt the equation of state $%
p_b=2n_ekT=\rho _b(2kT/m_p)=\rho _bc_b^2$, where $m_p$ is proton mass.

In general, the magnetic field will be non-uniform when the mode enters the
free-streaming regime. We note that, when the background fluid pressure was
large,the non-uniform magnetic field may have originally been part of an
Alfv\'en-type incompressible mode. However, as we mentioned earlier, once
this mode enters the free-streaming regime, there is a dramatic fall in the
fluid pressure, by a factor of order the very small baryon to photon ratio $%
\sim 10^{-9}$. As a result, the pressure of the non-uniform magnetic field,
associated with what might well have been an Alfv\'en-type mode, can no
longer be ignored, especially if the field exceeds $B_{crit}$. (Only a
perfectly circularly-polarised Alfv\'en wave has uniform magnetic pressure).
This non-uniform field associated with what started off as an Alfv\'en-type
mode, will now also induce gravitationally unstable, compressible motions.

In treating the resulting evolution, it is usual to assume (cf. Wassermann 
\cite{wasser} and Peebles \cite{peeb}) that perturbations to the Lorentz
force, due to that the perturbed velocity, are subdominant with respect to
the zeroth-order contribution of the Lorentz force itself. So one takes $%
{\bf B}={\bf B}_0({\bf x})a^2$, which solves the induction equation (\ref
{ind}), if ${\bf v}$ is neglected. Of course, this approximation will break
down once significant peculiar velocities have been developed, as will
always happen on sufficiently small scales, or at sufficiently late times,
for any given magnetic field. For galactic scales, it turns out that the
distortions to the magnetic field will become important only at late times,
even for $B_{-9}\sim 1$. So the above assumption of retaining only the
zeroth-order contribution to the Lorentz force is expected to be reasonable.
(The equations governing the more general case are derived in Appendix B,
and we hope to return to an analysis of this full system elsewhere). Making
these assumptions, standard linear perturbation analysis (cf. \cite{wasser}, 
\cite{peeb} , \cite{paddy}), leads to the evolution equation for $\delta _b$%
, 
\begin{equation}
{\frac{\partial ^2\delta _b}{\partial t^2}}+\left[ 2H+{\frac{4\rho _\gamma }{%
3\rho _b}}n_e\sigma _Ta\right] {\frac{\partial \delta _b}{\partial t}}-c_b^2{%
\nabla }^2\delta _b=4\pi Ga^2\left[ \rho _b\delta _b+\rho _c\delta _c\right]
+{\frac 1{a^3}}S_0({\bf x})  \label{delb}
\end{equation}
where the source term $S_0$ is given by 
\begin{equation}
S_0={\frac{{\bf \nabla }.\left[ {\bf B}_0\times ({\bf \nabla }\times {\bf B}%
_0)\right] }{4\pi \rho _b(t_0)}}.  \label{sorce}
\end{equation}
Here, $\rho _b(t_0)$ is the fluid density at the present time, $t_0$. If we
assume the dark matter to be cold, one can also derive a similar equation
for its fractional perturbed density $\delta _c$. One finds 
\begin{equation}
{\frac{\partial ^2\delta _c}{\partial t^2}}+2H{\frac{\partial \delta _b}{%
\partial t}}=4\pi Ga^2\left[ \rho _b\delta _b+\rho _c\delta _c\right] .
\label{delc}
\end{equation}

Prior to recombination, as pointed out in the previous section (Eq. (\ref
{scatexp}) ), the viscous damping dominates damping by the Hubble expansion.
For $B_0>B_{crit}$, the fluid pressure term can be neglected. Let us also
assume that the magnetic field was the only source of initial density
perturbations. Then, the compressible modes start their free-streaming
evolution with negligible initial $\delta _b$ since modes smaller than the
Silk scale have been significantly damped and modes on larger scales have a
negligible source of pressure perturbations in the radiation era because $%
\sim 3V_A^2/c^2\sim 3\times 10^{-7}B_{-9}$. Thus, Eq. (\ref{delb}) can be
solved, under the terminal velocity approximation, simply equating viscous
damping and the Lorentz force. We obtain 
\begin{equation}
{\frac{\partial \delta _b}{\partial t}}={\frac{S_0}{a^3}}({\frac{4\rho
_\gamma }{3\rho _b}}n_e\sigma _Ta)^{-1}=V_A^2{\frac{l_\gamma (t_0)}{l_B^2}}%
\equiv \gamma _d.  \label{tervel}
\end{equation}
Here, $l_B$ is a typical co-moving coherence scale over which the field
varies. Note that the RHS of this equation is constant in time and so the
density contrast $\delta _b$ increases linearly in this epoch. Hence, at the
time of recombination, $t_r$, the induced baryonic perturbation is $\delta
_b=\gamma _d(t_r-t_f)$, where $t_f$ is the time when the scale $%
k_p^{-1}(t)\sim l_Ba(t)$, becomes smaller than the photon mean-free-path.
For a flat universe, dominated by dark matter, the total fractional density
contrast in the matter is, $\delta =(\rho _b\delta _b+\rho _c\delta
_c)/(\rho _b+\rho _c)\sim \Omega _b\delta _b$. At the time of recombination,
we then have 
\begin{equation}
\delta (t_r)\approx 3.8\times 10^{-5}B_{-9}^2h^{-3}\left( {\frac{l_B}{1Mpc}}%
\right) ^{-2}(1-t_f/t_r).  \label{delbfr}
\end{equation}
On galactic scales with $l_B\sim 1Mpc$, we have $t_f/t_r\sim 0.4$, and then $%
\delta (t_r)\sim 2.1\times 10^{-5}h^{-3}B_{-9}^2$. This turns out to be
small compared to the $\delta $ induced in the post-recombination regime
(see below).

Next, consider the post-recombination evolution. The mean-free-path of the
photon now increases rapidly to a value exceeding the Hubble radius and
viscous damping becomes subdominant compared to expansion damping. Thus, we
can neglect the viscous damping term. Also, for $B >B_{crit}$, we can
neglect the fluid pressure term. Now suppose the baryons contribute a
fraction $f_B$ to the matter density while the cold dark matter contributes
a fraction $1-f_B$. Then, multiplying Eq. (\ref{delb}) by $f_B$, (\ref{delc}%
) by $1-f_B$, and adding the resulting equations, we get for the total
density contrast $\delta =(\rho _b\delta _b+\rho _c\delta _c)/(\rho_b +
\rho_c)$, 
\begin{equation}
{\frac{\partial ^2\delta }{\partial t^2}}+2H{\frac{\partial \delta }{%
\partial t}}-4\pi Ga^2\rho _m\delta ={\frac 1{a^3}}f_BS_0({\bf x}).
\label{del}
\end{equation}
Here, $\rho _m$ is the total matter density. Let us assume that at
recombination $\delta\approx 0$ and $(\partial \delta /\partial t)\approx 0$%
; that is, initially there are negligible fluctuations in density and
velocity divergence. Note that this is valid for scales much larger than
galactic scales. For scales with $l_B \sim Mpc$, it turns out that, the post
recombination evolution induces a $\delta$ much larger than that given in
Eq. (\ref{delbfr}), within an expansion time. So including this initial $%
\delta(t_r)$ gives negligible corrections. The particular solution of (\ref
{del}) for a flat matter-dominated universe is given by 
\begin{equation}
\delta (t)={\frac 9{10}}f_Bt_0^2S_0\left[ ({\frac t{t_r}})^{2/3}-{\frac 53}+{%
\frac 23}({\frac t{t_r}})^{-1}\right] .  \label{sol}
\end{equation}
This implies that the magnetic field induces a present-day fractional
density contrast $\delta _0=\delta (t_0)$, with 
\begin{equation}
\delta _0={\frac 9{10}}f_Bt_0^2S_0(1+z_r)\approx {\frac 9{10}}f_B\left( {%
\frac{V_{Ab}(t_0)t_0}{l_B}}\right) ^2(1+z_r)  \label{predel}
\end{equation}
where $(1+z_r)=(t/t_r)^{2/3}$ and $V_{Ab}(t_0)$ is the Alfv\'en velocity
with respect to the baryons given by 
\begin{equation}
{\frac{V_{Ab}(t_0)}c}={\frac{B_0}{\sqrt{4\pi \rho _b(t_0)c^2}}}\approx
1.9\times 10^{-5}B_{-9}f_b^{-1/2}  \label{baralf}
\end{equation}
Adopting $h=1/2$, $1+z_r=1100$, and a flat universe, with $t_0=2/(3H_0)$, $%
f_B=0.05$, we have 
\begin{equation}
\delta _0\approx 2.96B_{-9}^2({\frac{l_B}{1Mpc}})^{-2}.  \label{typdel}
\end{equation}
We see therefore that a magnetic field with $B_0\sim 10^{-9}G$ is needed to
impact significantly on galaxy formation. Such fields will also induce
rotational perturbations and give significant angular momentum to
protogalaxies \cite{wasser}.

>From Eq. (\ref{sol}) and (\ref{typdel}), it would seem that for
sufficiently small, $l_B=l_s$, say, one can have a $\delta _b\sim 1$, even
close to recombination. However, the calculation leading to Eq. (\ref{sol})
would break down on such small scales because the field distortions induced
by the motions, which we have neglected, will become dominant. The resulting
magnetic pressures will oppose gravity when the Alfv\'en crossing time
becomes of order the dynamical time; that is, for proper lengths $%
a(t)l_B<l_J(t)\sim V_{Ab}(t)t$, where $t$ is the relevant dynamical time
(the age of the universe) and $l_J(t)$ is the magnetic Jeans length. Noting
that $V_{Ab}(t)=V_{Ab}(t_0)a^{-1/2}(t)$ and $t=t_0a^{3/2}$ for a flat
universe, the comoving magnetic Jeans length is given by 
\begin{equation}
\lambda _J={\frac{l_J(t)}a}\sim V_{Ab}(t_0)t_0\sim 3.8\times
10^{-2}B_{-9}h^{-1}f_b^{-1/2}{\rm Mpc}.  \label{jean}
\end{equation}
Our treatment of how inhomogeneous magnetic fields induce structure
formation is valid only for $l_B>>\lambda _J$. On smaller scales one has to
solve the full set of equations outlined in Appendix B. We expect strong
magneto-sonic waves to be induced by such small scale inhomogeneities in a
sufficiently strong magnetic field. These may suffer strong dissipation and
so input energy into the IGM. We hope to return to this issue elsewhere.

\section{Discussion}

We have studied the evolution and damping of inhomogeneous magnetic fields
in various regimes. It is of interest to synthesise our results and discuss
how a given spectrum of magnetic inhomogeneities evolves. We make a few
general points and then describe the fate of magnetic inhomogeneities after
they enter the Hubble radius.

\begin{itemize}
\item  As we noted in Section IV, for magnetic field strengths $B_{-9}<1$,
the pressure perturbations are negligible. The fast compressible waves
induced by these perturbations, have a phase velocity of order the
relativistic sound speed, but negligible amplitudes with $\delta _b\sim
V_A^2/c_s^2\sim 3V_A^2/c^2$. The motions induced by the field can then be
treated as incompressible, at least until they enter the free-streaming
regime. These slow residual motions occur no faster than the Alfv\'en
timescale, and pressure can constantly readjust on the fast sound-crossing
time scale to preserve the incompressibility condition. In this case,
Alfv\'en modes, both non-linear and linear, and the incompressible limit of
the slow mode, that we studied in previous sections, are indeed the most
relevant.

\item  It is important to note that linearisation about a constant
background field is not a good approximation when following the evolution of
magnetic fields which are inhomogeneous with roughly similar power on a
multitude of scales. This was one of our motivations for concentrating on
the non-linear Alfv\'en mode, and studying its evolution and damping through
various epochs. Although this analysis employed special exact solutions, the
amplitude of the tangled component of the magnetic field could be taken to
be arbitrarily large compared to the amplitude of the large-scale field.
Also, its spatial configuration can be arbitrarily specified, by the free
function $b_0(\tau _0,z)$. So, one expects the behaviour of this mode to
reflect, at least qualitatively, the behaviour of general incompressible
motions driven by magnetic-field inhomogeneities.
\end{itemize}

A comoving scale $l_B$, which enters the Hubble radius in the
radiation-dominated epoch, does so at an epoch $t_e$, specified by $%
l_Ba(t_e)=2t_e$. Suppose we define this epoch by the radiation temperature $%
T(t)=T_0/a(t)$, where $T_0$ is the present-day microwave background
radiation temperature, then we have \cite{paddy} 
\begin{equation}
T(t_e)=T_e=63eV({\frac{l_B}{1Mpc}})^{-1}.  \label{entry}
\end{equation}
In models which produce the field during an inflationary epoch, the initial
condition for the non-linear Alfv\'en wave at the time of horizon entry
could be taken to be that the fluid is at rest but the field is tangled. The
Lorentz force due to this tangled field will then start pushing on the
fluid, when the scale of the tangle becomes smaller than the Hubble radius.
On the other hand, if the fields are produced in an early-universe phase
transition, they could be associated with large initial velocities.

\begin{itemize}
\item  In the cosmological context, it should be kept in mind that there is
only a finite time for the Alfv\'en wave to develop and induce motions in
the fluid, if they were initially absent. For example, on a comoving scale
of $k^{-1}$, by the time of recombination, at $\tau =\tau _r$, (or at a
temperature $T=T_r$), the Alfv\'en wave would have oscillated at most by a
phase angle of 
\begin{equation}
\chi =kV_A(\tau _r-\tau _e)<kV_A\tau _r\sim 1.8\times 10^{-1}B_{-9}({\frac{%
k^{-1}}{1Mpc}})^{-1}\left( {\frac{T_r}{0.25eV}}\right) ^{-1}.  \label{phase}
\end{equation}
Here, $\tau _e$ is the conformal time when a mode enters the Hubble radius,
and we have expressed $\chi $ in terms of the inverse of the wavenumber, $%
k^{-1}$. Thus, only small-scale magnetic inhomogeneities, with scales $%
k^{-1}<l_s\approx 0.1B_{-9}Mpc$, will have had time to oscillate by more
than $\pi /2$ in phase. Modes on scales $k^{-1}>>l_s$, which started with
zero initial peculiar velocity, and oscillated by a phase $\chi <<1$ when
damping is ignored, cannot damp their tangles, even if damping is included.
This situation applies to magnetic fields tangled on galactic scales, with $%
k^{-1}\sim Mpc$. Thus galactic-scale magnetic inhomogeneities, with $B_{-9}<1
$, do not get damped by photon viscosity, simply because the incompressible
wavelike motions they induce, oscillate negligibly before recombination.

\item  As the photon mean-free-path grows, the fast compressible motions,
induced by the magnetic field or due to existing ''adiabatic'' density
fluctuations, are damped on scales smaller than the Silk scale, that is on
scales $k^{-1}$ less than about $0.3L_S\sim 0.3(l_\gamma (t)t)^{1/2}$. The
comoving Silk damping scale, $L_S^C=L_S/a$, at any time $t$ (or temperature $%
T$) in the radiation era, is given by 
\begin{equation}
L_S^C(T)\approx 8.5\times 10^{25}\left( {\frac T{0.25eV}}\right)
^{-3/2}f_b^{-1/2}cm.  \label{silkco}
\end{equation}
After matter domination, when the scale factor $a(t)\propto t^{2/3}$, the
co-moving Silk scale is given by 
\begin{equation}
L_S^C(T)\approx 6\times 10^{25}\left( {\frac T{0.25eV}}\right)
^{-5/4}h^{-1/2}f_b^{-1/2}cm.  \label{silkcom}
\end{equation}

\item  By contrast, the evolution of non-linear Alfv\'en wave modes depends
on the ratio $D/\omega _0$. Non-linear Alfv\'en wave modes which enter the
Hubble radius when the temperature of the universe $T>2.5eV$, do so when $%
D/\omega _0<1$, and the mode is then initially in the damped oscillatory
regime. But the photon mean-free-path grows as $a^3$ while wavelengths grow
as $a$, and soon tangles on some scale are in the overdamped regime, with $%
D/\omega _0>1$. For a given scale, $k^{-1}$, this happens at a time $\tau
_{OD}$, when the temperature of the universe drops to $T<T_{OD}$, where 
\begin{equation}
T_{OD}=10.8eV\times B_{-9}^{-1/2}({\frac{k^{-1}}{1Mpc}})^{-1/2}f_b^{-1/2}.
\label{overdep}
\end{equation}
So, tangles on smaller scales not only enter the Hubble radius at an earlier
epoch, but are overdamped at an earlier epoch.

\item  While the Alfv\'en mode is in the damped oscillatory regime, its
evolution can be described by Eq. (\ref{oslim}), and so all the modes damp
by a factor of order $\exp (-\int^\tau (D/2))\sim \exp [-(3/4)(4k_p^2(\tau
)L_S^2(\tau )/45)]$. This is almost the same as the Silk damping factor, for
the usual baryon-photon sound waves. Hence, all modes which are smaller than
the Silk scale will get significantly damped by photon viscosity before the
non-linear Alfv\'en mode enters the overdamped regime. The largest comoving
scale, say $k^{-1}=k_{osc}^{-1}$, which gets damped while the mode is in the
damped oscillatory regime, can be estimated by equating $%
[k_{osc}L_S^C(T_{OD},k_{osc})]/15=1$. This gives 
\begin{equation}
k_{osc}^{-1}\approx 4.4\times 10^{-7}B_{-9}^3f_bMpc.  \label{oscdamp}
\end{equation}

\item  In the damped oscillatory regime, a mode which starts initially from
rest ($\dot f=0$), can be described by the solution $f(\tau )\approx f_0\cos
[\chi (\tau -\tau _e)]\exp [-\int (D/2)]$. Even modes which damp negligibly
in this above regime, will oscillate by a phase of order 
\begin{equation}
\chi (\tau _{OD})\sim kV_A\tau _{OD}\approx 4.14\times 10^{-3}B_{-9}^{3/2}({%
\frac{k^{-1}}{1Mpc}})^{-1/2}f_b^{1/2}  \label{odphase}
\end{equation}
when the mode enters the overdamped regime. Modes with $\chi (\tau _{OD})>1$
will acquire a large $\dot f$ by the time they enter the overdamped regime
and will be strongly damped. This will happen for any mode with $%
k^{-1}=k_{OD}^{-1}<1.7\times 10^{-5}B_{-9}^3f_b$ Mpc.

\item  It turns out that, for the Alfv\'en-type modes with $%
k^{-1}>5k_{OD}^{-1}$, which enter the Hubble radius with zero initial
velocity, there is negligible further damping while the mode is in the
overdamped regime. When these modes enter the overdamped regime, with $%
\omega _0/D=1$, we find that they do not have sufficient velocity ($\dot f$)
for friction to be important. With increasing time, the velocity grows, $D$
increases, and $\omega _0/D$ decreases. Eventually, the mode enters the
terminal-velocity regime, at say a time $\tau _T$, where friction balances
the Lorentz force (see Section IV). This happens roughly when $\tan (\chi
(\tau _T))\approx (\omega _0/D(\tau _T))<<1$ and $f(\tau _T)\sim f_0$. In
the terminal velocity regime, we have from Eq. (\ref{tersol}), that 
\begin{equation}
f(\tau )=f(\tau _T)\exp \left[ -\int_{\tau _T}^\tau {\frac{\omega _0^2}D}%
d\tau \right] \approx f(\tau _T)\exp \left[ -0.9\times 10^{-4}({\frac{T_T-T}{%
0.25eV}})B_{-9}^2\right] .  \label{termdam}
\end{equation}
For all modes with $l_B>5l_{OD}$, one finds that there is no significant
damping.

\item  When the photon mean-free-path increases above the wavelength of a
given mode, one enters the free-streaming regime. As we discussed in Section
VI, the further evolution of the Alfv\'en-type mode depends on the whether
magnetic pressure at this time dominates the fluid pressure, or vice versa.
This is determined by the ratio $B/B_{crit}$. For the case of $B<<B_{crit}$,
one can again treat the evolution as incompressible. The resulting Alfv\'en
modes are already in the overdamped regime when they start to free-stream
and their evolution is governed by Eq. (\ref{terovert}), 
\begin{equation}
\bar f(\tau )=\bar f(\tau _f)\exp -\left[ \int_{\tau _f}^\tau {\frac{\bar 
\omega _0^2}{\bar D}}dt\right] =\bar f(\tau _f)\exp -V_A^2\left[
\int_{t_f}^tk_p^2(t)l_\gamma (t)dt\right] .  \label{freedam}
\end{equation}
Hence, $\bar f(\tau )=\bar f(\tau _f)\exp (-k^2/k_{fs}^2)$, where the
free-streaming damping scale $k_{fs}^{-1}$ is given by 
\begin{equation}
k_{fs}^{-2}=V_A^2\int_{t_f}^t{\frac{l_\gamma (t)dt}{a^2(t)}}  \label{kfsdef}
\end{equation}
Modes with a scale for the magnetic field $k^{-1}<k_{fs}^{-1}$, get damped
significantly during the free streaming evolution. We see that the damping
in this regime is similar to Silk damping, except that the usual Silk
damping integral within the exponential (cf. Eq. (\ref{sldamp}) ) is
multiplied by an extra factor of $(15/2)V_A^2<<1$. For the linearised modes
where the wave vector makes an angle $\theta $ to the zero-order field, one
has to replace $V_A$ by $V_A\cos \theta $. After recombination, the viscous
damping is subdominant, compared to expansion damping (since $l_\gamma $
exceeds the Hubble radius), and so can be neglected. So the largest scale to
be damped is found by evaluating $k_{fs}^{-1}$ at the recombination
redshift. Assuming that the universe is matter dominated at recombination,
we get $k_{fs}^{-1}\approx (3/5)^{1/2}V_AL_S^C(t_r)$. Hence, the damping
scale is of order the Alfv\'en velocity times the Silk scale. The largest
wavelength mode to be damped, say $L_D^A\equiv 2\pi k_{fs}^{-1}(t_r)$ is
given by, 
\begin{equation}
L_D^A=2\pi ({\frac 35})^{1/2}V_A{\frac{L_S(t_r)}{a(t_r)}}\approx 1.1\times
10^{23}B_{-9}f_b^{-1/2}h^{-1/2}cm.  \label{lda}
\end{equation}

\item  For $B>B_{crit}$, we noted in Section VI that the evolution becomes
compressible, and gravitationally unstable for scales larger than the
magnetic Jeans length, $\lambda _J$. On such scales, we showed that a field
with $B_{-9}\sim 1$, is needed to produce a density perturbation large
enough to significantly affect galaxy formation. We did not treat the
evolution on scales smaller than the magnetic Jeans length in any detail,
although the governing equations are given in Appendix B (cf. Eq. (B8)). The
solution of these equations, as we noted in Section VII, is complicated by
the presence of an inhomogeneous zero-order magnetic field. Nevertheless, we
expect that fast compressible motions on scales smaller than $\lambda _J$
will drive oscillations close to the baryonic Alfv\'en frequency, and will
be initially overdamped by the action of photon viscosity, in the
pre-recombination era. The damping scale for such motions will then be
similar to $k_{fs}^{-1}$, as deduced above. For modes similar to the slow
magneto-sonic wave, we expect that for $B>>B_{crit}$, the phase velocity
will be roughly equal to $c_b$, the baryon sound speed, and the
corresponding damping scale will be $k_{fs}^{-1}$ with $V_A$ replaced by $c_b
$. These expectations are borne out by the linearised calculations of JKO,
although they ignore the inhomogeneous nature of the zero-order magnetic
field. (For a general tangled zero order field, the counterpart of the slow
wave may not be easy to excite). Clearly, more detailed computations are
needed to get the exact damping scales, in this case.

\item  One can consider the damping due to neutrino viscosity in the early
universe \cite{nu}, in an exactly analogous manner to the Silk damping
effects treated in detail above. We briefly describe below some of the
consequences of damping due to neutrinos. These damping effects are largest
around the time of neutrino decoupling, at a temperature of $T=T_\nu \sim
1MeV$. During this epoch, the number density of weakly interacting particles
is $n_W\sim T^3$ (cf.\cite{kolb},\cite{paddy}). The cross-section for
interaction with neutrinos, is typically the weak interaction cross-section $%
\sigma _W\sim G_F^2T^2$,where $G_F$ is the Fermi constant. The neutrino
mean-free-path is then given by 
\begin{equation}
l_\nu ={\frac 1{\sigma _Wn_W}}\approx 1.4\times 10^{11}cm\left( {\frac T{MeV}%
}\right) ^{-5}.  \label{numfp}
\end{equation}
Note that $l_\nu $ becomes comparable to the Hubble radius, $%
H^{-1}=2t\approx 4.4\times 10^{10}cm(T/MeV)^{-2}$, at $T\sim 1MeV$. The
comoving damping scale for fast modes, say $L_S^{C\nu }$, due to neutrino
viscosity, can be derived in exactly analogous manner to the Silk damping
scale, derived in Section V. We get $L_S^{C\nu }=[t(T_\nu )l_\nu (T_\nu
)]^{1/2}/a(T_\nu )\sim l_\nu (T_\nu )/a(T_\nu )\approx 0.9\times 10^{20}cm$.
As with the damping by photon viscosity, the maximum damping of Alfv\'en (or
the slow) modes occur when they enter the free-streaming regime. We analyse
the damping of the non-linear Alfv\'en wave in the neutrino free-streaming
regime, in Appendix C. We show there that damping in this regime is similar
to Silk damping except that the usual Silk damping integral within the
exponential (cf. Eq. (\ref{sldamp}) ) is multiplied by an extra factor of $%
(15/2)V_{A\nu }^2<<1$, and $l_\gamma $ is replaced by $l_\nu $. Here, $%
V_{A\nu }$ is the Alfv\'en velocity defined in terms of the conserved,
neutrino energy density $\rho _\nu (T_\nu )a^4(T_\nu )$. Since $l_\nu
\propto T^{-5}\propto t^{5/2}$ in the radiation-dominated epoch, we can
estimate the largest wavelength Alfv\'en mode that is appreciably damped, by
neutrino viscosity, in the free streaming regime. This comoving wavelength
is given by $L_{D\nu }^A\sim 2\pi (2/5)^{1/2}V_{A\nu }L_S^{C\nu }\sim
10^{17}B_{-9}cm$.
\end{itemize}

\noindent We can generalise these results given any initial spectrum of
magnetic inhomogeneities, by replacing $V_A$ by a suitably averaged
scale-dependent $V_A(k^{-1})$.

\section{Conclusions}

We have considered the evolution and viscous damping of cosmic magnetic
fields in the early universe in detail. Using the fact that the fluid,
electromagnetic, and shear viscous energy-momentum tensors are all
conformally invariant, we showed in Section II that the MHD equations in the
FRW universe, including viscous effects, can be transformed into their
special-relativistic counterpart when the metric perturbations from
inhomogeneous motions are small. This enabled us to transform known
non-linear Alfv\'en-wave solutions, from flat spacetime, into the expanding
FRW universe.

We considered in detail the evolution and damping of these modes in various
regimes. First, on galactic scales or larger, the Alfv\'en mode oscillates
negligibly before recombination, for magnetic field strengths, $B_{-9}<1$
(or a present day magnetic field $B<10^{-9}$ $Gauss$). So there is then no
question of strong ``Silk'' damping of these modes, due to photon viscosity,
as occurs for compressional baryon-photon oscillations. Furthermore,
Alfv\'en waves with small enough wavelength, which can oscillate appreciably
before recombination, become overdamped. In this case, the longest
wavelength which suffers appreciable damping by photon viscosity, has a
scale $k^{-1}\sim V_AL_S^C$, where $L_S^C$ is the usual comoving Silk scale.
Since the Alfv\'en speed is $V_A\sim 3.8\times 10^{-4}B_{-9}<<1$, only
comoving wavelengths smaller than $10^{23}B_{-9}$ $cm$ suffer appreciable
damping. We also briefly considered analogous results for damping of very
small scales by neutrino viscosity following neutrino decoupling at $t\sim 1s
$.

After recombination, the fluid pressure drops enormously, roughly by the
baryon to photon ratio. The Lorentz force due to a tangled magnetic field
associated with an initially Alfv\'en-like mode before recombination can
seed gravitationally unstable compressional perturbations after
recombination, provided the field is strong enough and tangled enough on
scales larger than the magnetic Jeans length, $\lambda _J$. We examined the
post-recombination evolution of scales larger than $\lambda _J$ in section
VII, including the effect of a passive dark matter component, and showed
that magnetic fields with $B_{-9}\sim 1$ are needed to impact significantly
on galaxy formation. The evolution equations for perturbations with scales
smaller than $\lambda _J$ are derived in Appendix B, but their solution is
much more complicated and will be examined elsewhere.

Magnetic fields with $B_{-9}\sim 1$, may be constrained by observations of
quasar rotation measures \cite{kronberg}. The fluid velocities induced by
the tangled field, oscillating as an Alfv\'en mode in the pre-recombination
era, can also produce small angular scale anisotropies in the microwave
background, through the Doppler effect \cite{ksjdb}. This may provide
another constraint on such fields. The dissipation of magnetic fields due to
neutrino or photon viscosity will also leave an imprint on the spectrum of
neutrinos and photons respectively. The neutrino spectrum could be probed
using nucleosynthesis and there already exist strong limits on the spectral
distortions of the microwave background. Scenarios of galaxy formation that
appeal to strong enough magnetic fields of order $10^{-9}$ $Gauss$ are
potentially testable. We hope to return to a further consideration of some
of these observational issues in future work.

\acknowledgements

KS was supported by a PPARC Visiting Fellowship at the Astronomy Centre,
University of Sussex. He thanks the staff there for warm hospitality. JDB is
supported by a PPARC Senior Fellowship.

\appendix

\section{Maxwell's equation in the "Lab" frame}

In the main text we defined the electric field ${\bf E}^{*}\equiv
(E^{*1},E^{*2},E^{*3})$ and magnetic field ${\bf B}^{*}\equiv
(B^{*1},B^{*2},B^{*3})$ in the starred metric by 
\begin{equation}
F^{*0i}=E^{*i}\quad F^{*12}=B^{*3}\quad F^{*23}=B^{*1}\quad F^{*31}=B^{*2}.
\label{elbdefa}
\end{equation}
This can be written in more compact way in term of a four-vector electric
field $E_\mu ^{*}$ and magnetic field $B_\mu ^{*}$ as 
\begin{equation}
B_\mu ^{*}={\frac 12}\epsilon _{\mu \nu \rho \lambda }^{*}V^{*\nu }F^{*\rho
\lambda };\quad E_\mu ^{*}=F_{\mu \nu }^{*}V^{*\nu },  \label{ebnudef}
\end{equation}
where $V^{*\mu }\equiv [1,0,0,0]$ is the four-velocity of fundamental
observers at rest in the starred metric frame. We have also used the
Levi-Civita tensor $\epsilon _{\mu \nu \rho \lambda }^{*}=\sqrt{-g^{*}}{\cal %
A}_{\mu \nu \rho \lambda }$, with ${\cal A}_{\mu \nu \rho \lambda }$, the
totally antisymmetric symbol such that ${\cal A}_{0123}=1$ and $\pm 1$ for
any even or odd permutations of $(0,1,2,3)$. Note that the four-vectors $%
B_\mu ^{*}$ and $E_\mu ^{*}$ have purely spatial components and $%
E_i^{*}=E^{*i},B_i^{*}=B^{*i}$ for the spatial components of the field
4-vector. First, let us transform the electric and magnetic four-vectors to
the unstarred (FRW) conformal frame, and denote the resulting electric and
magnetic four-vectors by $E_\mu ^{\prime }$ and $B_\mu ^{\prime }$,
respectively. Making use of the conformal transformation properties, 
\begin{equation}
F^{*\mu \nu }=\Omega ^{-4}F^{\mu \nu };\quad \epsilon _{\mu \nu \rho \lambda
}^{*}=\Omega ^4\epsilon _{\mu \nu \rho \lambda };\quad V^{*\nu }=\Omega
^{-1}V^\nu ,  \label{contr}
\end{equation}
we get 
\begin{equation}
B_\mu ^{\prime }=\Omega B_\mu ^{*}={\frac 1a}B_\mu ^{*};\quad E_\mu ^{\prime
}=\Omega E_\mu ^{*}={\frac 1a}E_\mu ^{*}.  \label{preb}
\end{equation}
We now make a coordinate transformation to the FRW proper ''Lab''
co-ordinates $(t,{\bf r})$ defined by $dt=ad\tau $, $d{\bf r}=ad{\bf x}$.
Denote the co-ordinate-transformed, ''Lab'' electric and magnetic
four-vectors by $E_\mu $ and $B_\mu $, respectively. We have 
\begin{equation}
B_\mu ={\frac 1a}B_\mu ^{\prime }={\frac 1{a^2}}B_\mu ^{*};\quad E_\mu ={%
\frac 1a}E_\mu ^{\prime }={\frac 1{a^2}}E_\mu ^{*}.  \label{labeb}
\end{equation}
Note that these 4-vectors are also purely spatial. Similarly, we can define
the current density in the Lab frame by $J_L^\mu =aJ^\mu =a\Omega ^4J^{*\mu
}=a^{-3}J^{*\mu }$. Now, denote the spatial components $B_\mu $, $E_\mu $
and $J_L^\mu $ by the spatial 3-vectors ${\bf B}$, ${\bf E}$ and ${\bf J}$
respectively. Then, in terms of these spatial vectors, we have 
\begin{equation}
{\bf B}={\frac{{\bf B^{*}}}{a^2}};\quad {\bf E}={\frac{{\bf E^{*}}}{a^2}}%
;\quad {\bf J}={\frac{{\bf J^{*}}}{a^3}}  \label{spatq}
\end{equation}
The four-current density will also have a time component, the charge density 
$\rho _q=J_L^0=(J^{*0}/a^3)$. Using the Maxwell equations, (\ref{maxstar}),
in the starred metric, one can write the Maxwell equations in terms of these
''Lab'' fields. We have 
\begin{equation}
{\bf \nabla }\times (a^2{\bf B})=4\pi a^3{\bf J}+{\frac{\partial (a^2{\bf E})%
}{\partial \tau }};\quad {\bf \nabla }.{\bf B}=0  \label{maxlaba}
\end{equation}
\begin{equation}
{\bf \nabla }\times (a^2{\bf E})=-{\frac{\partial (a^2{\bf B})}{\partial
\tau }};\quad {\bf \nabla }.(a^2{\bf E})=4\pi a^3\rho _q.  \label{maxlabb}
\end{equation}
The non-relativistic limit of Ohm's law in terms of the ''Lab'' fields is
simply 
\begin{equation}
{\bf E}+{\bf v}\times {\bf B}={\frac{{\bf J}}\sigma }.  \label{ohmlab}
\end{equation}
In the ideal limit, the ''Lab'' magnetic field then satisfies the induction
equation 
\begin{equation}
{\frac{\partial (a^2{\bf B})}{\partial t}}={\frac 1a}{\frac{\partial (a^2%
{\bf B})}{\partial \tau }}={\frac 1a}{\bf \nabla }\times \left[ {\bf v}%
\times (a^2{\bf B})\right] .  \label{indap}
\end{equation}
So, when ${\bf v}=0$, we have ${\bf B}\propto a^{-2}$, a result which is
intuitively expected for the ''Lab'' magnetic field due to flux freezing in
the expanding universe. It is also of interest to express the Euler equation
(\ref{euler}) for the fluid in the radiation-dominated era, in terms of the
''Lab'' fields. We obtain 
\begin{eqnarray}
{\frac 1{a^4}}{\frac \partial {\partial t}}\left[ (\rho +p)a^4{\bf v}\right]
+{\frac{({\bf v}.{\bf \nabla })}a}\left[ (\rho +p){\bf v}\right] &+&{\frac{%
{\bf v}}a}{\bf \nabla }.\left[ (\rho +p){\bf v}\right]  \nonumber \\
\ &=&-{\frac 1a}{\bf \nabla }p+{\bf J}\times {\bf B}+{\frac \eta {a^2}}%
\left[ \nabla ^2{\bf v}+{\frac 13}{\bf \nabla }({\bf \nabla }.{\bf v}%
)\right] .  \label{eulerlab}
\end{eqnarray}

\section{Generalised linear perturbation theory in the free-streaming and
post-recombination regimes}

Begin with the linearised evolution equations for the baryonic fluid,
including the effects of the magnetic field and gravity. We have 
\begin{equation}
{\frac {\partial {\bf v}}{\partial t }} + H(t){\bf v} = -{\frac{1}{a \rho_b}}%
{\bf \nabla }p_1 + {\frac{1 }{\rho_b}} {\bf J}\times {\bf B} - {\frac{1}{a}}%
{\bf \nabla}\phi - {\frac{4\rho_{\gamma} }{3\rho_b}} n_e \sigma_T {\bf v}
\label{eulerina}
\end{equation}
\begin{equation}
{\frac{\partial \delta_b}{\partial t }}+{\frac{1 }{a}}{\bf \nabla }.{\bf v}
=0.  \label{conta}
\end{equation}
\begin{equation}
{\bf \nabla}^2\phi = 4\pi G a^2 \delta\rho_T = 4\pi G a^2
\left[\rho_b\delta_b + \rho_c\delta_c\right]  \label{poisona}
\end{equation}
Here $p_1= c_b^2 \rho_b \delta_b$ is the perturbed pressure. We take ${\bf B}
= [{\bf B}_0({\bf x}) + {\bf b}({\bf x})]/a^2$, with $\vert {\bf b} \vert <<
\vert {\bf B}_0 \vert$. The linearised induction equation, (\ref{indap}),
then becomes 
\begin{equation}
{\frac{\partial {\bf b}}{\partial t }} = {\frac{1}{a}}{\bf \nabla }\times
\left[ {\bf v}\times {\bf B}_0\right] ; \quad {\bf \nabla}.{\bf b}= 0 .
\label{inda}
\end{equation}

Suppose the perturbation is described in terms of the perturbed position of
the baryonic component by $\delta {\bf x}_b={\bf \xi }({\bf x},\tau )$. The
perturbed velocity is ${\bf v}=a(\partial {\bf \xi }/\partial t)$. An
integration of the perturbed continuity equation (\ref{conta}) and induction
equation (\ref{inda}) then gives 
\begin{equation}
\delta _b=-{\bf \nabla }.{\bf \xi }\;\quad {\bf b}={\bf \nabla }\times
\left[ {\bf \xi }\times {\bf B}_0\right] .  \label{rhobla}
\end{equation}
Similarly, one can define the perturbed comoving position of the cold dark
matter component by $\delta {\bf x}_c={\bf \xi }_c({\bf x},\tau )$. An
integration of the perturbed continuity equation for this component gives $%
\delta _c=-{\bf \nabla }.{\bf \xi }_c$. The Poisson equation (\ref{poisona})
for the potential can then be integrated once to give 
\begin{equation}
-{\frac 1a}{\bf \nabla }\phi =4\pi Ga\left[ \rho _b{\bf \xi }+\rho _c{\bf %
\xi }_c\right] .  \label{poi}
\end{equation}
The perturbed pressure-gradient term can be written as 
\begin{equation}
-{\frac 1{a\rho _b}}{\bf \nabla }p_b={\frac{c_b^2}a}{\bf \nabla }({\bf %
\nabla }.{\bf \xi }).  \label{pri}
\end{equation}
Using Eq. (\ref{rhobla}), (\ref{poi}) and (\ref{pri}), the perturbed Euler
equation becomes 
\begin{eqnarray}
{\frac{\partial ^2{\bf \xi }}{\partial t^2}}+\left[ 2H+{\frac{4\rho _\gamma 
}{3\rho _b}}n_e\sigma _T\right] {\frac{\partial {\bf \xi }}{\partial t}}={%
\frac{c_b^2}{a^2}}{\bf \nabla }({\bf \nabla }.{\bf \xi }) &+&4\pi Ga\left[
\rho _b{\bf \xi }+\rho _c{\bf \xi }_c\right] +{\frac{({\bf \nabla }\times 
{\bf B}_0)\times {\bf B}_0}{(4\pi \rho _ba^3)a^3}}  \nonumber \\
&&+{\frac{\left[ {\bf \nabla }\times ({\bf \nabla }\times [{\bf \xi }\times 
{\bf B}_0])\right] \times {\bf B}_0}{(4\pi \rho _ba^3)a^3}}  \nonumber \\
&&+{\frac{({\bf \nabla }\times {\bf B}_0)\times \left[ ({\bf \nabla }\times [%
{\bf \xi }\times {\bf B}_0)\right] }{(4\pi \rho _ba^3)a^3}.}  \label{fineq}
\end{eqnarray}
This linear equation describes the gravitationally unstable evolution. and
damping of linearised MHD modes in the free-streaming and post-recombination
regime. However, one cannot perform a simple Fourier analysis of this
equation, since ${\bf B}_0$ is also a function of ${\bf x}$. In the main
text we have simplified the equation by neglecting the last two terms. While
this approximation is likely to be valid for large-scale modes at early
times, it will break down once the distortion of the magnetic field due to
the motions become significant. We hope to return to the study of this
equation elsewhere.

\section{ Viscous damping due to neutrinos in the free-streaming regime}

Modes whose wavelengths become longer than the neutrino mean-free-path,
enter the neutrino free-streaming regime. The evolution and damping of
non-linear Alfv\'en wave solutions can be examined in this regime, analogous
to the case of photon-free streaming, treated in section VI. First the
viscous force due to coupling of the relativistic plasma and the neutrinos
is given by\cite{nu} 
\begin{equation}
{\bf F}_D^\nu =-{\frac 43}n_W\sigma _W\rho _\nu {\bf v}={\frac 43}\rho _\nu {%
\frac{{\bf v}}{l_\nu }}.  \label{nufric}
\end{equation}
Here, $\rho _\nu $ is the energy density of neutrinos, and we have replaced $%
(n_W\sigma _W)^{-1}$ by the neutrino mean-free-path $l_\nu $. We can also
use the form of the Euler and induction equations, (as derived in Appendix
A); except that the viscous force is as given by Eq. (\ref{nufric}) and the
fluid inertia/pressure does not include the neutrino contributions.
Following the ideas of Section IV and VI, we look again at non-linear
Alfv\'en modes with ${\bf B}=({\bf B}_0+{\bf b})/a^2$, where ${\bf B}_0=B_0%
\hat {{\bf z}}$, with $B_0={\rm constant}$, ${\bf b}={\bf n}g(\tau )e^{ikz}$
and ${\bf v}={\bf n}\bar v_g(\tau )e^{ikz}$, with ${\bf n}$ perpendicular to 
$\hat {{\bf z}}$. Recall that $|{\bf b}|$ is {\it not necessarily small}
compared to $|{\bf B}_0|$. Also, since the universe is still radiation
dominated at the epochs near neutrino decoupling, with $p>>B^2/(8\pi )$, we
can treat the motions as nearly incompressible. The Euler equation gives 
\begin{equation}
{\frac{4\rho _Ra^4}3}{\frac{\partial {\bf v}}{\partial \tau }}={\frac{B_0}{%
4\pi }}{\frac{\partial {\bf b}}{\partial z}}-{\frac 43}\rho _\nu a{\frac{%
{\bf v}}{l_\nu }}.  \label{eulnu}
\end{equation}
where $\rho _R$ is the fluid density, excluding the neutrino contribution.
The induction equation gives 
\begin{equation}
{\frac{\partial {\bf b}}{\partial \tau }}=B_0{\frac{\partial {\bf v}}{%
\partial z}}  \label{indnu}
\end{equation}
These can be combined to give an equation for the evolution of $g(\tau )$ 
\begin{equation}
{\frac{d^2g}{d\tau ^2}}+D_\nu {\frac{dg}{d\tau }}+\omega _\nu ^2g=0,
\label{barfnu}
\end{equation}
where 
\begin{equation}
\omega _\nu =kV_{AR};\quad D_\nu ={\frac a{l_\nu }}({\frac{4\rho _\nu }{%
3\rho _R}})\propto \tau ^{-4}.  \label{nudecdef}
\end{equation}
Here, $V_{AR}^2=3B_0^2/(16\pi \rho _Ra^4)={\rm constant}$, is the Alfv\'en
velocity defined with the inertia contributed by $\rho _R$.

The evolution of this non-linear Alfv\'en mode depends once again on the
relative strengths of the damping and driving terms. When modes enter the
neutrino free-streaming regime, the Alfv\'en waves are again strongly
overdamped, with $D_\nu /\omega _\nu >>1$, for $B_{-9}<1$. One can again
apply the terminal-velocity approximation of Section IV in the overdamped
regime, once $dg/d\tau $ has adjusted itself to the ''zero acceleration''
solution. Then, $g$ is given by 
\begin{equation}
g(\tau )=g(\tau _i)\exp \left( -\int_{\tau _i}^\tau k^2V_{A\nu }^2{\frac{%
l_\nu }a}d\tau \right) ,  \label{ternu}
\end{equation}
where 
\begin{equation}
V_{A\nu }^2={\frac{B_0^2}{4\pi (4\rho _\nu a^4/3)}}  \label{nualf}
\end{equation}
is the Alfv\'en velocity defined in terms of the conserved neutrino energy
density $\rho _\nu a^4$. We see that the damping of these Alfv\'en modes by
neutrino viscosity is similar to Silk damping except that the usual Silk
damping integral within the exponential (cf. Eq. (\ref{sldamp})) is
multiplied by an extra factor of $(15/2)V_{A\nu }^2<<1$, and $l_\gamma $ is
replaced by the neutrino mean-free-path $l_\nu $.\\

\bigskip 

*Presently at: Max Plank Institute for Astrophysics, Karl Schwarzschild
Strasse 1, 85748 Garching, Germany. On leave from National Centre for Radio
Astrophysics, TIFR Pune, Poona University Campus, Ganeshkhind, Pune 411 007,
India.\\


\bigskip 

\begin{references}
\bibitem{dynam}  Moffat, H. K., {\it Magnetic Field Generation in
Electrically Conducting Fluids}, Cambridge University Press, Cambridge,
(1978); Krause, F. $\&$ Radler, K.-H., {\it Mean-Field Magnetohydrodynamics
and Dynamo Theory}, Pergamon Press, Oxford, (1980); Ruzmaikin, A. A.,
Shukurov, A. M. $\&$ Sokoloff, D. D., {\it Magnetic Fields of Galaxies},
Kluwer, Dordrecht (1988); Mestel, L., {\it Stellar Magnetism}, Oxford UP,
Oxford, (1998); Beck, R., Brandenburg, A., Moss, D., Shukurov, A. $\&$
Sokoloff, D., Ann. Rev. Astron. Astrophys., {\bf 34}, 155 (1996).

\bibitem{seedrev}  Vishniac, E.T., Ap. J. {\bf 446}, 724 (1995); Rees, M.
J., Quart. Jl. Roy. Astron. Soc. {\bf 28}, 197 (1987); Rees, M.J., In {\it %
Cosmical Magnetism}, ed. Lynden-Bell, D., Kluwer, London, (1994) p155;
Subramanian, K., Bull. Astr. Soc. Ind., {\bf 23}, 481 (1995); Subramanian,
K., Narasimha, D. $\&$ Chitre, S.M., Mon. Not. R. astron. Soc. {\bf 271},
L15 (1994), and references therein. 

\bibitem{euseed}  Hogan, C. J., Phys. Rev. Lett., {\bf 51}, 1488 (1983);
Turner, M. S. $\&$ Widrow, L. M., Phys. Rev. D, {\bf 30}, 2743 (1988);
Vachaspati, T., Phys. Lett. B, {\bf 265,} 258 (1991). Ratra, B.,. Ap.J
Lett., {\bf 391}, L1 (1992); Gasperini, M., Giovannini, M. $\&$ Veneziano,
G., Phys. Rev. Lett. {\bf 75,} 3796 (1995); Lemoine, D., $\&$ Lemoine, M.,
Phys. Rev. D, {\bf 52}, 1955 (1995); O. T\"ornkvist, hep-ph/9707513; K.
Enqvist, astro-ph/9707300; Giovaninni, M., and Shaposhnikov, M.E.,
hep-ph/9710234; Joyce, M. and Shaposhnikov, M.E., Phys. Rev. Lett., {\bf 79}%
, 1193 (1997).

\bibitem{dyndeb}  Cattaneo, F. $\&$ Vainshtein, S. I., Ap.J., {\bf 376}, L21
(1991); Vainshtein, S. $\&$ Rosner, R., Ap.J., {\bf 376}, 199 (1991);
Kulsrud, R.M. $\&$ Anderson, S.W., Ap.J., {\bf 396}, 606 (1992);
Brandenburg, A., In {\it Lectures on Solar and Planetary Dynamos}, ed.
Proctor, M. R. E. $\&$ Gilbert, A. D., Cambridge University Press, (1994),
p117; Field, G. B., {\it The Physics of the Interstellar Medium and
Intergalactic Medium}, ed. Ferrara {et al.}, ASP Conf. Ser., {\bf 80,}
(1996), p1; Blackman, E., Phys. Rev. Lett.,{\bf \ 77}, 2694 (1996);
Chandran, B., Princeton University Observatory preprint (1996) POPe-659;
Cattaneo, F. $\&$ Hughes, D. W., Phys. Rev E,{\bf \ 54}, 4532 (1996);
Subramanian, K., Mon. Not. R. astron. Soc. in press; astro-ph/9707280.

\bibitem{primhyp}  Kulsrud, R., IAU Symp. 140: {\it Galactic and
Extragalactic Magnetic Fields}, Reidel, Dordrecht, (1990), p527, and
references therein.

\bibitem{barrow}  J.D. Barrow, P.G. Ferreira, and J. Silk, Phys. Rev. Lett. 
{\bf 78}, 3610 (1997); J.D. Barrow, Phys. Rev. D {\bf 55}, 7451 (1997).

\bibitem{silk}  Silk, J., Ap. J., {\bf 151}, 431 (1968).

\bibitem{JKO96}  Jedamzik, K., Katalinic, V., and Olinto, A., (1996),
astro-ph/9606080 v2

\bibitem{weinb}  Weinberg, S., {\it Gravitation and Cosmology}, Wiley, New
York, (1972); G.F.R. Ellis, in {\it General Relativity and Cosmology, }ed.
R.K. Sachs, Varenna Lectures in Physics, Academic, New York, (1971).

\bibitem{brand}  Brandenburg, A., Enqvist, K., Olesen, P., Phys. Rev. D, 
{\bf 54}, 1291 (1996).

\bibitem{laing}  Liang, E. P. T., Ap. J., {\bf 211}, 361 (1977).

\bibitem{parker79}  Parker, E. N., {\it Cosmic Magnetic Fields}, Clarendon
Press, Oxford (1979).

\bibitem{ll}  Landau, L. $\&$ Lifshitz, E. M., {\it Electrodynamics of
Continuous Media}, Pergamon Press, Oxford (1987).

\bibitem{walker}  Walker, T. P., Steigman, G., Schramm, D. N., Olive, K. A. $%
\&$ Kang, H.-S., Ap.J., 376, {\bf 51} (1991).

\bibitem{hata}  Steigman, G., Hata, N. $\&$ Felten, J. E., astro-ph/9708016
and references therein.

\bibitem{wasser}  Wasserman, I., Ap.J., {\bf 224}, 337 (1978)

\bibitem{peeb}  Peebles, P. J. E., {\it Large Scale Structure of the Universe%
}, Princeton University Press, Princeton N. J., (1980).

\bibitem{paddy}  Padmanabhan, T., {\it Structure Formation in the Universe},
Cambridge University Press, Cambridge (1993).

\bibitem{kolb}  Kolb, E. W. $\&$ Turner, M. S., {\it The Early Universe },
Addison-Wesley, New York, (1990) .

\bibitem{kaiser}  Kaiser, N.,, Mon. Not. R. astron. Soc., {\bf 202}, 1169
(1983).

\bibitem{kronberg}  Kronberg, P. P., Rep. Prog. Phys., {\bf 57}, 325 (1994)

\bibitem{ksjdb}  Subramanian, K. $\&$ Barrow, J. D., 1998 (in preparation).

\bibitem{nu}  Neutrino viscosity was first considered as a means of damping
anisotropies and inhomogeneities by C.W. Misner, Nature {\bf 214}, 40 (1967)
and Phys. Rev. Lett. {\bf 19}, 533 (1967). See also, J.D. Barrow, Nature 
{\bf 267}, 117 (1977). The damping of very small scale irregularities by GUT
scale transport processes is discussed by J.D. Barrow, Mon. Not. R. astron.
Soc. {\bf 199,} 45P (1982).
\end{references}
\end{document}